\documentclass{rsproca_new}%%%%where rsproca is the template name
\usepackage[numbers, authoryear]{natbib}
\usepackage{xspace}
\usepackage{float}
\usepackage{standalone}
\usepackage{tikz}
\usepackage{graphicx}
\usepackage{pgf}
\usepackage{intcalc}
\usepackage{enumitem}
\usepackage{mathrsfs}  
\usepackage{orcidlink}
\usepackage{upgreek}

\setlist[description]{leftmargin=\parindent,labelindent=\parindent}

\usepackage{acro}
\acsetup{
  single    = false,
  list/sort  = true,
  cite/group = true,
  cite/group/cmd = \citealp,
  cite/group/pre = {;\xspace},
}

\DeclareAcronym{gpu}{
  short = GPU,
  long = Graphics Processing Unit
}

\DeclareAcronym{ode}{
  short = ODE,
  long = Ordinary Differential Equation
}

\DeclareAcronym{cbm}{
  short = CBM,
  long = Carbon Box Model,
  cite = {Craig1957,oeschger1975,Siegenthaler1980,Nakamura1987,dorman2004}
}
\DeclareAcronym{mcmc}{
  short = MCMC,
  long = Markov Chain Monte Carlo,
  cite = {Metropolis1953}
}

\DeclareAcronym{scpm}{ 
  short = SCP-M,
  long = Simple Carbon Project,
  cite = {ONeill2019}
}

\DeclareAcronym{spe}{
  short = SPE,
  long = Solar Proton Event
}
\DeclareAcronym{gp}{
  short = GP,
  long = Gaussian Process,
  cite = {rasmussen:williams:2006}
}

\DeclareAcronym{snr}{
  short = SNR,
  long = signal-to-noise ratio
}

\usetikzlibrary{arrows.meta,backgrounds,fit,positioning}
\usetikzlibrary{external}

\usepackage{lipsum}  
\newcommand{\dfc}{\mbox{$\Updelta^{14}\text{C}$}\xspace}
\newcommand{\fc}{\mbox{$^{14}\text{C}$}\xspace}
\newcommand{\tbe}{\mbox{$^{10}\text{Be}$}\xspace}
\newcommand{\tcl}{\mbox{$^{36}\text{Cl}$}\xspace}
\newcommand{\ce}{\,CE\xspace}
\newcommand{\bce}{\,BCE\xspace}
\newcommand{\ticktack}{\texttt{ticktack}\xspace}

\begin{document}

%%%% Article title to be placed here
\title{Modelling Cosmic Radiation Events in the Tree-ring Radiocarbon Record}

\author{%%%% Author details - add your middle names and any alternate names here
Qingyuan Zhang$^{1\,\orcidlink{0000-0002-0906-8533}}$, Utkarsh Sharma$^{1\,\orcidlink{0000-0002-0771-8109}}$, Jordan A. Dennis$^{1\,\orcidlink{0000-0001-8125-6494}}$, Andrea Scifo$^{2\,\orcidlink{0000-0002-7174-3966}}$,\\ Margot Kuitems$^{2\,\orcidlink{0000-0002-8803-2650}}$,  Ulf B\"{u}ntgen$^{3,4,5,6\,\orcidlink{0000-0002-3821-0818}}$,\\ Mathew J. Owens$^{7\,\orcidlink{0000-0003-2061-2453}}$, Michael W. Dee$^{2\,\orcidlink{0000-0002-3116-453X}}$, \\ and Benjamin J. S. Pope$^{1,8\, \orcidlink{0000-0003-2595-9114}}$}

%%%%%%%%% Insert author address here
\address{\footnotesize {$^{1}$School of Mathematics and Physics, University of Queensland, St Lucia, QLD 4072, Australia\\
$^{2}$Centre for Isotope Research, University of Groningen, Groningen, the Netherlands\\
$^3$Department of Geography, University of Cambridge, Cambridge, CB2 3EN, UK\\
$^4$Global Change Research Institute (CzechGlobe), Czech Academy of Sciences, 603 00, Brno, Czech Republic\\
$^5$Department of Geography, Faculty of Science, Masaryk University, 611 37, Brno, Czech Republic\\
$^6$Swiss Federal Research Institute (WSL), 8903, Birmensdorf, Switzerland\\
$^{7}$Department of Meteorology, University of Reading, Earley Gate, PO Box 243, Reading RG6 6BB, UK\\
$^{8}$Centre for Astrophysics, University of Southern Queensland, West Street, Toowoomba, QLD 4350, Australia}
}

%%%% Subject entries to be placed here %%%%
\subject{computer modelling and simulation, astrophysics, atmospheric science}

%%%% Keyword entries to be placed here %%%%

%%%% Insert corresponding author and its email address}
\corres{Benjamin J. S. Pope\\
\email{b.pope@uq.edu.au}}

%%%% Abstract text to be placed here %%%%%%%%%%%%
\begin{abstract}
% miyake events are mysterious
Annually-resolved measurements of the radiocarbon content in tree-rings have revealed rare sharp rises in carbon-14 production. These `Miyake events' are likely produced by rare increases in cosmic radiation from the Sun or other energetic astrophysical sources.
% need to model global carbon cycle
The radiocarbon produced is not only circulated through the Earth’s atmosphere and oceans, but also absorbed by the biosphere and locked in the annual growth rings of trees. To interpret high-resolution tree-ring radiocarbon measurements therefore necessitates modelling the entire global carbon cycle.
% ticktack is an open source model that talks to bayesian inference engines
Here, we introduce ‘\href{https://github.com/SharmaLlama/ticktack/}{\ticktack}’, the first open-source Python package that connects box models of the carbon cycle with modern Bayesian inference tools.
% we infer parameters using mcmc and find interesting results
We use this to analyse all public annual \fc tree data, and infer posterior parameters for all six known Miyake events. They do not show a consistent relationship to the solar cycle, and several display extended durations that challenge either astrophysical or geophysical models. 
% we encourage people to adapt our methods etc
% Extending these methods to modelling stable carbon isotopes, or the radionuclides $^{36}$Cl and $^{10}$Be, we
\end{abstract}
%%%%%%%%%%%%%%%%%%%%%%%%%%%

%%%%%%%%%% Insert the texts which can accommodate on first page in the tag "fmtext" %%%%%

\begin{fmtext}

\end{fmtext}

%%%%%%%%%%%%%%% End of first page %%%%%%%%%%%%%%%%%%%%%

\maketitle

\section{Introduction}
% introduce why we have radiocarbon dating 
% libby 1949 and early history (kern2020)
Radiocarbon dating is used to accurately determine the age of samples of biological material, and is a fundamental tool of modern archaeology \citep{kern2020,Hajdas2021}.
% radiocarbon production is from thermal neutrons
Thermal neutrons produced by cosmic rays interact with $^{14}$N atoms in the upper atmosphere to produce radioactive \fc, or radiocarbon, which filters across the carbon cycle through the atmosphere, biosphere, and marine environments. \citet{libby49} demonstrated that the ratio of \fc to stable carbon isotope abundances is approximately constant in the atmosphere over time: while living organisms continually replenish \fc from the atmosphere, in dead organic matter this radiocarbon decays with a 5700-year half life, and therefore can be used as a clock to date archaeological and palaeontological samples.

% calibration curve: production rate varies with time. suess1970bristle to IntCal20 (intcal13, Reimer2020, Heaton2020, Hogg2020)
In detail, this picture is complicated by variations in the radiocarbon production rate. The most relevant source of variation in the context of this work is from the activity cycle of the Sun. At low points in solar activity, reduced magnetic shielding means that the cosmic ray flux at Earth is increased \citep{Stuiver1993}; but also, shocks ahead of solar coronal mass ejections can accelerate energetic particles that produce radiocarbon in Earth's atmosphere. As a result, radiocarbon measurements are not only important tools for archaeology, but also for historical studies of space weather, solar and geomagnetic activity, and the Earth's climate dynamics \citep{Heaton2021}.

For many species, tree-rings can be dated to the exact year of their formation, the science of dendrochronology. Radiocarbon in tree-rings, appropriately adjusted for radioactive decay, therefore offers a detailed record of radiocarbon concentrations over time. The existence of variation from one year to the next was first shown by \citet{deVries1958}. Using measurements on North American bristlecone pine, \citet{suess1970bristle} revealed the scale of radiocarbon fluctuations over millennial time scales, demonstrating the necessity for a `calibration curve` for archaeological dating. Such curves have attained increasing sophistication over time, and have in the last decade attained high precision and annual resolution, for example IntCal13 \citep{intcal13} and IntCal20 \citep{Reimer2020,Heaton2020,Hogg2020}.

% miyake 2012: first miyake event in 774\,\ce
These newly detailed curves revealed the long-suspected astrophysical influence of the solar activity cycle on modulating radiocarbon production in individual solar cycles \citep{guttler13}. They also yielded a surprise: \citet{miyake12} discovered in Japanese cedar tree-rings a sudden single-year jump in radiocarbon concentration around 774\ce. 
% other miyake events - 993 AD (miyake 2013), 660 BCE (park2017), 5259 BCE (Brehm 21), 5410 BCE (Miyake et al 2021), 7176 BCE (Brehm 21),
This was followed shortly by the discovery of another spike in tree-rings from 993\ce \citep{miyake13}, and further such spikes have been found in 660\bce \citep{park2017}, 5259\bce \citep{Brehm2022_events}, 5410\bce \citep{miyake21}, and 7176 BCE \citep{Brehm2022_events}, for a total of six well-studied and accepted radiocarbon spikes. These are often known as `Miyake events', after their first discoverer.
% false positive in 3372–3371 BCE (Wang et al, 2017, Jull et al, 2021), 800 BC (jull2018). AD ones are globally coherent (buntgen 2018), duration of 660 is prolonged (sakurai 20)
Other spikes have been claimed from some tree-ring samples, but not replicated globally: one in 800\bce \citep{jull2018}, and claimed but refuted in 3372\bce \citep{Wang2017,jull2021}. Several small events are also proposed in 1052\ce, and 1261, 1268, and 1279\ce by \citet{brehm21} and \citet{Miyahara2022}.

Detailed study of these events is important to determine their origin. Better data are available for the two events in the Common Era, showing that the events of 774 and 993 CE are globally coherent, including many trees in both the Northern and Southern Hemispheres \citep{buntgen18}. Meanwhile, although the other events show sharp single-year rises, the event of 660\bce has a prolonged rise over a couple of years, which could be due to a prolonged production or a succession of events \citep{sakurai2020}. For comparison, a decade-long rise in 5480\bce, less than a century before the single-year rise in 5410\bce, is ascribed by multiradionuclide evidence to an unusual grand solar minimum of very great depth and short duration \citep{miyake17,kanzawa2021}. No other sharp rises in \dfc so far detected have shown evidence of substructure in time.

% archaeological use - dee16_rspa, changbaishan (oppenheimer2017,hakozaki2018), por-bajin (kuitems2020), l'anse aux meadows (Kuitems2021), viking age Ribe (Philippsen2021), etc.
Miyake events offer archaeologists a sharp radiocarbon signal, synchronized across the Earth, which can be used to achieve single-year dates for tree-rings in samples otherwise beyond the reach of dendrochronology \citep{dee16_rspa}. For example, the historically-significant eruption of the Changbaishan volcano can be dated to 946\,\ce \citep{oppenheimer2017,hakozaki2018}. By dating the Uyghur site of Por-Bajin in Russia to exactly 777\,\ce, it can be identified as a monastery built under the Uyghur Khaganate's short-lived conversion to Manichaeism \citep{kuitems2020}. These data have been most revolutionary for Viking Age archaeology. The 774\,\ce event dates finds at Ribe, Denmark, and anchors interpretation of their trade networks \citep{Philippsen2021}, while the 993\,\ce event securely dates the L'Anse aux Meadows settlement to 1021\,\ce\,-- the first evidence of European settlement in the Americas \citep{Kuitems2021}. 

The sharp rise in radiation, with a simultaneous global onset, indicates that Miyake events are of astrophysical origin, for which a variety of explanations have been offered \citep[thoroughly  reviewed by][]{Cliver2022}.
% explanations: magnetars (wang 2019)? supernovae? (dee16_radiocarbon) 
Dying stars and their remnants are known to produce extremely intense bursts of radiation, and are \textit{prima facie} reasonable astrophysical sources. For instance, a sharp burst of radiation could have been delivered by a Galactic gamma-ray burst \citep{Hambaryan2013,pavlov2013} or nearby supernova, though astronomical evidence of these is so far lacking. \citet{dee16_radiocarbon} have failed to find evidence of a radiocarbon rise associated with any of the known historical supernovae, while \citet{Terrasi2020} find a 2$\sigma$ increase in radiocarbon in 1055\,\ce after the Crab supernova. An alternative proposal considers a magnetar burst from a nearby magnetized neutron star \citep{wang2019}, which is energetically plausible - but no sufficiently nearby or active neutron star is yet known from conventional astronomical observations. \citet{Pavlov2019a} and \citet{Pavlov2019b} have suggested prolonged events like 660\,BCE and 5480\,BCE are the result of enhanced Galactic cosmic ray flux over several years after the heliosphere is compressed by dense clouds in the interstellar medium.  Closer to home, \citet{Liu2014} suggest the \fc could be deposited into the atmosphere directly by a passing comet; this interpretation is rejected by \citet{usoskin2015_comet}, who argue that such a comet would need to have been of a size ($\gtrapprox 100$\,km) that would have devastated the Earth.

% best candidate is solar - usoskin, melott, mekhaldi papers
The wide consensus of the literature is that these events have a solar origin, beginning with \citet{melott2012,usoskin2013}. For example, the events could represent a solar magnetic collapse, a very brief grand solar minimum, with the reduced heliospheric shielding exposing the Earth to an increase in Galactic cosmic rays \citep{Neuhauser2015b}. Alternatively, and more popular in the literature, the Miyake events could represent the extreme tail of a distribution of solar flares continuous with those that are observed astrophysically on the modern Sun and other solar-like stars. We are fortunate that \fc is not the only cosmogenic isotope that can trace these events: we see evidence of the 774\ce and 993\ce events in time series of \tbe and \tcl from ice cores \citep{miyake15_be,mekhaldi2015}, and because the production of these isotopes depends on input particle energy, they can be used to infer a particle energy spectrum similar to solar energetic protons \citep{Webber2007}. Only the most energetic particles produce \tbe, but \tcl is expected to be produced at comparatively low energies and may therefore shed light on other events as well \citep{mekhaldi2021}. Extreme solar flares or emissions are plausible astrophysically: based on the findings of the \textit{Kepler} Space Telescope \citep{Borucki2010}, G~dwarf stars (like the Sun) are thought to produce superflares every few hundred to few thousand years \citep{okamoto2021}, even  old and slowly-rotating stars \citep{nogami2014,notsu2019}. 

Nevertheless, even in light of the uncertainties in particle flux from the existing literature, an event like the 774\ce event would need to be more than an order of magnitude larger than even the Carrington event, the most significant coronal mass ejection and accompanying geomagnetic storm ever observed in the instrumental era of science \citep{hudson2021}. By considering possible beaming angles and uncertainties in models of the carbon cycle, \citet{neuhauser2014} argue that the 774\ce event might be implausibly huge to be a single solar superflare. The solar proton event of 1956 produced an estimated $3.04\times10^6$ atoms/cm$^{-2}$ of \fc \citep{usoskin2020}; depending on assumptions about its flare class and spectral hardness, the 774\ce event could correspond to an X-ray flare as bright as X1800, nearly two orders of magnitude larger than any previously observed \citep{Cliver2020}.

Meanwhile, ice core nitrate records at 774\ce and 993\ce do not show any hint of a signal from extreme solar activity \citep{mekhaldi2017,Sukhodolov2017}. At least some superflares observed from other stars are known in fact to originate from unresolved M~dwarf binary companions \citep{jackman2021}, which are much more active than G~dwarfs like the Sun and, because we do not have such a companion ourselves, could not explain the radiocarbon bursts. 
% relation to solar cycle miyake 2013b, park2017, scifo2019, brehm2021, 
Extreme geomagnetic storms preferentially occur around the maxima of the solar cycle \citep{Owens2021}. While the historical data on solar energetic particle events is far more limited, it is reasonable to assume they follow a similar pattern \citep{barnard2018}, as both result from energetic coronal mass ejections. Thus if Miyake events occur preferentially at solar maxima, this would support a solar origin. 
The radiocarbon data themselves contain the 11-year solar cycle, and several attempts have been made to determine its phase at the time of a Miyake event \citep{miyake13b,FogtmannSchulz2017,park2017,scifo19,brehm21}, and in this Paper we will attempt a similar inference.  % add in some Kepler superflare literature

Unfortunately, there is fairly limited evidence in written historical accounts for unusual astronomical phenomena coinciding with the radiocarbon spikes \citep[for a comprehensive account, see][]{stephenson2015}. The Anglo-Saxon Chronicle reports a ``red crucifix, after sunset'' in 774\ce \citep{Allen2012}; if this is an aurora, this is consistent with a massive solar flare, but it has been argued that the `crucifix' is simply a lunar optical halo \citep[an interpretation rejected by \citealp{hayakawa2019}]{neuhauser2015}. An aurora is also reported in 775\,\ce from the Chinese chronicle Jiutangshu \citep{Zhou2014}. It remains the case that other historical records have not conclusively been shown to refer to aurorae in the year around this event.

% interest in long term solar cycle: failure of gyrochronology angus2015 explained by weakened magnetic breaking vansaders2016 and hall2021. idea of two solar cycles bv2007, metcalfe2016. superflares like carrington hudson2021
Understanding the long-term behaviour of solar activity is of current interest in astrophysics. A grand minimum in stellar activity has only been observed in one star other than the Sun \citep{Baum2022,luhn22}, and the Sun's own dynamo may be unusual. Solar-like stars are born rapidly rotating and very magnetically active, and their magnetized winds slow their rotation as they age -- so that the age of solar-like stars might be inferred from appropriately calibrated relations of `gyrochronology' \citep{barnes2003}. No single gyrochronology relation, however, fits the rotation periods of large samples of stars determined with \textit{Kepler} \citep{angus2015}. The emerging consensus is that weakened magnetic braking in older stars causes the activity to diminish without commensurate reduction in rotation periods \citep{vansaders2016,hall2021,metcalfe2022}, and this may be caused by a transition occurring at Rossby numbers of order unity between a fast and a slow type of stellar magnetic dynamo \citep{bohmvitense2007,metcalfe2016}. Remarkably, not only is the Sun less active than most solar-like stars \citep{Reinhold2020}, it so happens that our own Sun is at about the age and Rossby number of the proposed transition - so that it may be atypical of field stars generally, and long term time series of its activity are of broad relevance in astrophysics. 

If a Miyake event were to occur today, the sudden and dramatic rise in cosmic radiation could be devastating to the biosphere and technological society. It is therefore concerning that we have little understanding of how to predict their occurrence or effects. A solar proton event orders of magnitude more powerful than any previously observed could cause an `internet apocalypse' of prolonged outages by damaging submarine cables and satellites \citep{Jyothi2021}. The direct effects of energetic particles could even harm the health of passengers in high-altitude aircraft \citep{fujita2021,hubert2021,Sato2008}. It is also likely that the 774\ce event would have caused a $\sim 8.5\%$ depletion in global ozone coverage, with a significant but not catastrophic effect on weather \citep{Sukhodolov2017}. The origin and physics of these radiocarbon spikes are therefore important not just for astronomers and archaeologists, but for risk planning and mitigation in general society.

\subsection{Carbon Cycle Models}

% have to model the entire global carbon cycle - see craig et al 1957, dorman 2004, bomb peak nakamura 1987
A very short pulse of radiation striking the atmosphere leads to a sharp rise  ($\sim 1$\,yr) in measured \dfc and slow decay ($\sim$ decade timescale) as the new radiocarbon is filtered through the global carbon cycle, findings its way into the biosphere, oceans, and sediments. Therefore to interpret radiocarbon time series astrophysically, it is necessary to model this carbon cycle. The most popular way of doing this is using a \ac{cbm}, in which the global carbon budget is partitioned between discrete reservoirs (e.g. the atmosphere, oceans, and biota, or subdivisions thereof). It is also common to include effects of atmospheric circulation or geochemistry in other areas of geoscience and planetary science \citep[e.g.][]{KrissansenTotton2017,ONeill2019}, but on the timescales and sensitivities relevant to Miyake events these reservoirs are assumed to be coupled to one another linearly. This leads to a system of first order ordinary differential equations (ODEs): a diffusion process, with an inhomogeneous driving term for atmospheric production.

% previous models: guttler15, miyake17, buntgen18, brehm21
While \ac{cbm}s are essential for relating tree-ring time series to production rates, none of those models applied to Miyake events in the literature are available open-source. As a consequence, different analyses contain model-dependent systematic effects that are hard to reproduce or calibrate. 

% our approach - open source and reproducible
In this Paper, we introduce a fast Python framework for carbon box models, \href{https://sharmallama.github.io/ticktack}{ticktack}\footnote{Named for the Malvina Reynolds song, \textit{Little Boxes} (1962), in which little boxes are \textit{all made out of ticky-tacky / And they all look just the same}.}. The framework is designed to be flexible, allowing arbitrary box models to be specified and modified. This is implemented in the high-performance Google \textsc{Jax} library \citep{jax}, which supports just-in-time compilation, automatic differentiation, and code deployment to \ac{gpu}s. This code interfaces with the popular Bayesian inference packages \texttt{emcee} \citep{emcee} and \textsc{JaxNS} \citep{jaxns}.
% infer parameters for all events - amplitude, duration
We use this to reproduce several recent \ac{cbm}s applied to radiocarbon time series: the 4-box model of \citet{miyake17}, the 11-box model of \citet{guttler15}, and the 22-box models of \citet{buntgen18} and \citet{brehm21}. 

We apply these to all published annual tree-ring data on all six known Miyake events, and infer posterior probability distributions for parametric and nonparametric models of the radiocarbon production rate over time, including the timing and duration, amplitude, and relation to the solar cycle. These posteriors determine a relationship to the solar cycle in 993\ce, 774\ce, and 5410\bce, though not for other events, and a range of total radiocarbon production delivering in a single pulse the equivalent of 1 -- 4 years of average production. 

\section{Methods: the ticktack Carbon Box Model Framework} 
\label{sec:methods}

% previous models are closed source
Carbon box models are widely used in literature from archaeology to geophysics. They span different levels of sophistication, from simple treatments of radiocarbon relative to carbon-12, through to full models of global geochemistry since the beginning of the Earth \citep[e.g.][]{KrissansenTotton2017}. 

On the timescales that are relevant to single-year spikes of radiation, it is sufficient to consider only the dynamics of radiocarbon against a fixed background of equilibrium carbon flows. The overall properties of these models are specified by the reservoirs into which carbon is partitioned; the stable carbon content of each reservoir $N^{12}_i$, and the stable carbon flows specified in Gt/yr or in residence times (yr), $F^{12}_{ij}$; the reservoirs in which radiocarbon is produced by cosmic rays, and in what proportions, $V_i$; and the long-term average production rate of radiocarbon $q_0$. 

The radiocarbon flux between reservoirs is then computed as

\begin{equation}
    F^{14}_ij = \underbrace{(\frac{m_{14}}{m_{12}N^{12}_i} \,F^{12}_{ij} - \lambda)}_{\equiv M_{ij}} \cdot N^{14}_i
\end{equation}

\noindent where $\lambda$ is the radioactive decay constant for \fc, and $M_{ij}$ is a static transfer matrix. This allows us to simplify the \ac{cbm} model for a radiocarbon state vector $\mathbf{y} \equiv [N^{14}_i]$ and vector of production coefficients $\mathbf{V}\equiv [V_i]$ as a linear, first order \ac{ode} 

\begin{equation}
\label{eq:cbm_ode}
    \frac{d \mathbf{y}}{dt} = \mathbf{M}\mathbf{y} + Q(t) \mathbf{V}
\end{equation}

\noindent where the inhomogeneous term $Q(t)$ is the radiocarbon production rate. For constant $Q(t) = q_0$ this has a steady state solution $\mathbf{y}_0 = \mathbf{M}^{-1} q_0$. For computational reasons, we reparameterize the \ac{ode} to the form 

\begin{equation}
    \frac{d(\mathbf{y}-\mathbf{y}_0)}{dt} = (Q(t)-q_0)\mathbf{V}
\end{equation}

\noindent which can be efficiently solved with a range of adaptive step-size algorithms. 
The results also depend to some extent on assumptions made in matching model outputs to data, including the growth seasons of trees and any short-term atmospheric dynamics; and in fitting these models to data, the algorithms used for optimization and inference.

% our model is open source and reproducible
We have developed an open-source, object-oriented Python package, \ticktack, for specifying and running arbitrary \ac{cbm}s. A user can input a series of \texttt{Box} and \texttt{Flow} objects with a numerical value for the reservoir or flow, units, and metadata (eg northern or southern hemisphere, or the fraction of radiocarbon production in this box) and then compiles this to a \texttt{CarbonBoxModel} object; or they can load a pre-saved object. The user can then specify an equilibrium production condition - either directly a radiocarbon production rate, or it can find the production rate by gradient descent to reach a target \fc quantity in a particular reservoir. 

This \texttt{CarbonBoxModel} then has a method \texttt{run} which uses the \textsc{Jax} Dormand-Prince \citep[DP5;][]{dp5} algorithm as implemented in the Diffrax differential equation library \citep{kidger2021} to solve the \ac{cbm} \ac{ode} for a specified initial condition, production rate, and timesteps. Because this is implemented in Google \textsc{Jax} \citep{jax}, this can be compiled, executed on GPUs, and is automatically differentiable, allowing for use in gradient descent optimization and Hamiltonian Monte Carlo \citep{betancourt2017}. 

% implements existing models
We have followed the descriptions of four models used in the literature which are sufficiently well-described in terms of carbon reservoirs and flows to be emulated in \ticktack, and which have been applied to Miyake event analysis: the 11-box \citet{guttler15} and 4-box \citet[Figure~\ref{fig:miyake-guttler}]{miyake17}, and 22-box \citet[Figure~\ref{fig:buntgen_model}]{buntgen18}, and \citet[Figure~\ref{fig:brehm_model}]{brehm21}. The 22-box models represent similar, but slightly different, partitions into 2 hemispheres of the global carbon cycle described in the 11-box model. All four models are available as default pre-saved models in \ticktack.

% Not a replacement for detailed models of the climate cycle
% this could go in the JOSS paper
\ticktack is not a replacement for detailed models of the climate cycle, but rather for fast reconstruction of production from tree-ring data. Open source alternatives such as \texttt{pyhector} \citep{pyhector}, Pymagicc \citep{pymagicc}, or the \ac{scpm} are geared towards climate modelling, for which our model is not sufficiently accurate, but are not fast enough to couple to Bayesian inference of radiocarbon production. \ac{scpm} couples ocean dynamics to a carbon cycle model, with $\sim 30$\,s runtime for 10\,ky; we need to achieve $\ll 1\,\text{s}$ runtime for \ac{mcmc}.
% Also not a replacement for radiocarbon date calibration software
We also do not aim to perform radiocarbon date calibration, for which there are several open-source libraries already available such as OxCal \citep{ramsey1995radiocarbon,ramsey2013recent}, BCal \citep{bcal}, MatCal \citep{matcal}, or ChronoModel \citep{chronomodel}.

\subsection{Parametric Inference}
\label{sec:parametric_inference}

% ties to Bayesian inference tools - SingleFitter / MultiFitter
We can use this model in forwards-mode to simulate time series of \fc or \dfc; and therefore also to solve the inverse problem of reconstructing radiocarbon production rates from data. We can load a tree-ring \dfc time series together with a \ac{cbm} as a \texttt{SingleFitter} class object in \ticktack, which has methods for parametric and nonparametric Bayesian inference of the production rate, or by direct inversion of the \ac{ode}. 
In this Paper, we adopt a parametric model for production rate $Q(t)$, given steady state $q_0$, including three components: 

\begin{equation}
    Q(t) = q_0 + A_\odot\,q_0 \sin{(\frac{2\pi t}{11\,\text{yr}} + \phi)} + S(t, t_0,\Delta t) + m\cdot t
\end{equation}

\noindent where the solar cycle has an amplitude $A_\odot$ and phase $\phi$; there is a long-term trend with gradient $m$. The Miyake event spike profile $S(t)$ is represented as a normalized super-Gaussian with start date $t_0$, duration $\Delta t$, and amplitude $S_0$:
\begin{equation}
    S(t, t_0,\Delta t) \equiv S_0/\Delta t \exp{(- \frac{t- (t_0 + \Delta t/2)}{1/1.93516\,\Delta t}^{16})} 
\end{equation}

\noindent The super-Gaussian form is chosen to approximate a top-hat function, but with differentiable sides more amenable to optimizers and \ac{ode} solvers. The numerical factor of 1.93516 is the integral of the unit super-Gaussian $\exp{(-t^{16})}$, and is used for normalization. The amplitudes of any of these coefficients can optionally be fixed at zero to disable each component of the production model.

This forwards model can be used with Bayesian tools to infer posterior probability distributions over the values of any of these parameters. We assume a Gaussian distribution for each \dfc sample with mean $d_i$  and  uncertainty $\sigma_i$, so that the log-likelihood of a parameter vector $\theta$ is

\begin{equation}
    \log \mathcal{L}(\theta) = \sum_i \, \frac{d_i - Q(\theta)_i}{\sigma_i}
\end{equation}

\noindent and adopt uniform priors over phase and start date with reasonable limits, and log-uniform Jeffreys priors over all other parameters.

The \texttt{SingleFitter} class has methods for sampling from this posterior using \ac{mcmc} as implemented in the affine-invariant ensemble sampler \texttt{emcee} \citep{goodman2010,emcee}, and nested sampling as implemented in \textsc{JaxNS} \citep{skilling2004,jaxns}. 

\subsection{Nonparametric Inference}
It is also possible to infer radiocarbon production rates per year directly from a \dfc time series, using either a direct inverse to the \ac{ode}, or by a forwards model with a flexible, high-dimensional parametrization. 

The ODE can be solved exactly for a box $i$ with nonzero production and measured data, such as from a tree-ring time series in the troposphere, by rearranging Equation~\ref{eq:cbm_ode} to the form
 
\begin{equation}
\label{eq:inverse_solver}
    Q = \dfrac{\dot{y_i} - (\mathbf{M}\mathbf{y})_i}{V_i},
\end{equation}

\noindent except that the flow term $\mathbf{M}\mathbf{y}$ depends on the radiocarbon state in \emph{all} boxes simultaneously, so that it is also necessary to infer the missing components of the state vector.

% brehm inverse solver
To implement this inverse solver, \citet{brehm21} take annual-cadence data, interpolate this to a continuous fine grid (for example to 12-month sampling), and in a finite-difference form of the \ac{cbm} \ac{ode} to iteratively find the production rate at each timestep to reach the required tropospheric \dfc measurement at the next time step.

In \ticktack we implement an alternative non-iterative approach, by interpolating $y_i(t)$ linearly, and using \textsc{Jax} to differentiate this to obtain $\dot{y}_i (t)$. We can then obtain the full state history $\mathbf{y}(t)$ by solving the \ac{ode} with the production term from Equation~\ref{eq:inverse_solver} and an initial steady state $\mathbf{y}_0$, and then use this completed state history to obtain $Q(t)$. This is exact for a finely sampled completed model, and in practice is a good approximation for data binned over a growth season if the time stamps are taken to be the middle of each growth season.

Because this inverse solver method relies on differentiation, when the \ac{snr} is low it has the tendency to amplify noise on short timescales. 
% gaussian process
In order to find a reconstruction that is more tolerant to noise, we want to use a Bayesian method as described above in Section~\ref{sec:parametric_inference}, but choose a very flexible high-dimensional parameterization for $Q(t)$. Here, we will use a set of control points - a large but finite grid of points $\mathbf{q} \equiv [q(t_i)]$ - as parameters, and use a Matérn--$3/2$ \ac{gp} both to interpolate these to a smooth function of time, and also use the \ac{gp} likelihood to penalize spurious short-timescale variations. We implement this \ac{gp} calculation using the \texttt{tinygp} library \citep{tinygp}, which is written in \textsc{Jax} and can therefore be compiled and differentiated along with the rest of \ticktack. 

% fourier alternative
While we do not attempt to do so here, it is also possible to solve this problem in the Fourier domain. The impulse response function of the carbon cycle to a pulse of radiation can be analytically determined as the matrix exponential

\begin{equation}
    \textbf{g}(t) = \mathbf{V} \exp{(-\mathbf{M}\,t)}
\end{equation}

\noindent and an arbitrary time series in box $i$ generated by the convolution $g_i(t) \star Q(t)$. The Fourier transform of $g_i(t)$ is a frequency response function that can be used as a linear filter in the Fourier domain, which \citet{usoskin2005} use as an alternative to iterative solution, but which is not implemented in \ticktack in this study.

\subsection{Tree-Ring Data}
\label{sec:data}

We apply this code to an analysis of all publicly-available \dfc data for the six events previously identified in the literature, gathering the COSMIC network data from many sites across both hemispheres for 774\ce and 993\ce from \citet{buntgen18}; additional data, including early and late wood data, for 774\ce from \citet{Uusitalo2018}, and Danish oak over 993\,\ce from \citet{FogtmannSchulz2017}; English oak over 993\,\ce from \citet{rakowski_2018}; the discovery data for 7176\bce and 5259\bce from \citet{Brehm2022_events}; earl tree-rings over 660\bce from \citet{park2017}, and early and late wood over 660\bce from \citet{sakurai2020}; and data from the decades leading up to 5410\bce from \citet{miyake17}. We exclude the Japanese cedar from \citet{Miyake2022}, as it shows a delayed rise compared to other 993\,\ce datasets, and for the purposes of the present work we await a consensus on how to interpret this.

% problems with 775 rise 
Before examining the modeled outputs in detail, some general observations should be made about the reliability and sensitivity of the underlying data. The apparent congruence of the sets of \dfc results, in both timing and amplitude, is especially remarkable given the data come from trees of various genera and species that grew in wide range of different habitats. In reality, every individual tree is subject to its own specific environment and the biotic and abiotic disturbances that it poses, such as insect outbreaks, fungal diseases and climate anomalies \citep{buntgen18}. As well as this, the physiology of each species determines the way it uses and/or reuses carbohydrates for growth-ring construction. This latter consideration lies at the core of an ongoing debate about whether whole rings or only late wood fractions should be analyzed to achieve the highest quality data \citep{FogtmannSchulz2017, mcdonald2019,Park2021}. Furthermore, at per mille precisions intra-annual fluctuations in atmospheric radiocarbon concentrations also become significant, and in particular the coincidence of annual maxima and minima with the growing seasons of different species at different locations \citep{kromer2001}. Finally, each laboratory employs its own celluose extraction technique, sometimes tailoring it to the individual species at hand. Such methodological differences have been shown to produce variations in data quality, even on samples of the same tree-rings \citep{Wacker2020}. 

Given all these complications, it is unsurprising that differences can be seen across the full suite of radiocarbon profiles. Nonetheless, there is one particular pattern which defies simple explanation, and may have some as yet unknown physical origin. Some 774\ce data sets exhibit an instantaneous uplift between 774 and 775~CE, while others show a more gradual rise over several years. The split between such sharp and prolonged rises in \dfc exists between different trees from similar environments, and even between trees of the same species from similar environments, as discussed in Section~\ref{sec:discussion}\ref{sec:timing_774}. It is unclear whether this effect is astrophysical, environmental, to do with unknown tree-growth dynamics, or a systematic in the measurements.
% \begin{itemize}
%     \item Data from many sites across both hemispheres for 774\ce and 993\ce from \citet{buntgen18}
%     \item Additional data, including early and late wood data, for 774\ce from \citet{Uusitalo2018}
%     \item Discovery data for 7176\bce and 5259\bce from \citet{Brehm2022_events}
%     \item Data for 660\bce from \citet{park2017}, and early and late wood for 660\bce from \citet{sakurai2020}
%     \item Data from the decades leading up to 5410\bce from \citet{miyake17}
% \end{itemize}

\section{Results}

\subsection{Parametric Fits}
\label{sec:parametric_fits}

We used the workflow scheduling package Snakemake \citep{snakemake} to automatically execute and reproducibly log parametric fits as described in Section~\ref{sec:methods}\ref{sec:parametric_inference} to all six events, with the 774\,\ce event split into sharp and prolonged rise subsets. While we had the option to apply nested sampling, we used the affine-invariant ensemble MCMC sampler \texttt{emcee} exclusively in this Section. 

We infer the start data, duration, spike amplitude, phase and amplitude of the solar cycle, and a long-term linear trend, with the model initialized in steady state with the solar cycle.

\begin{figure}
    \centering
    \includegraphics[scale=0.43]{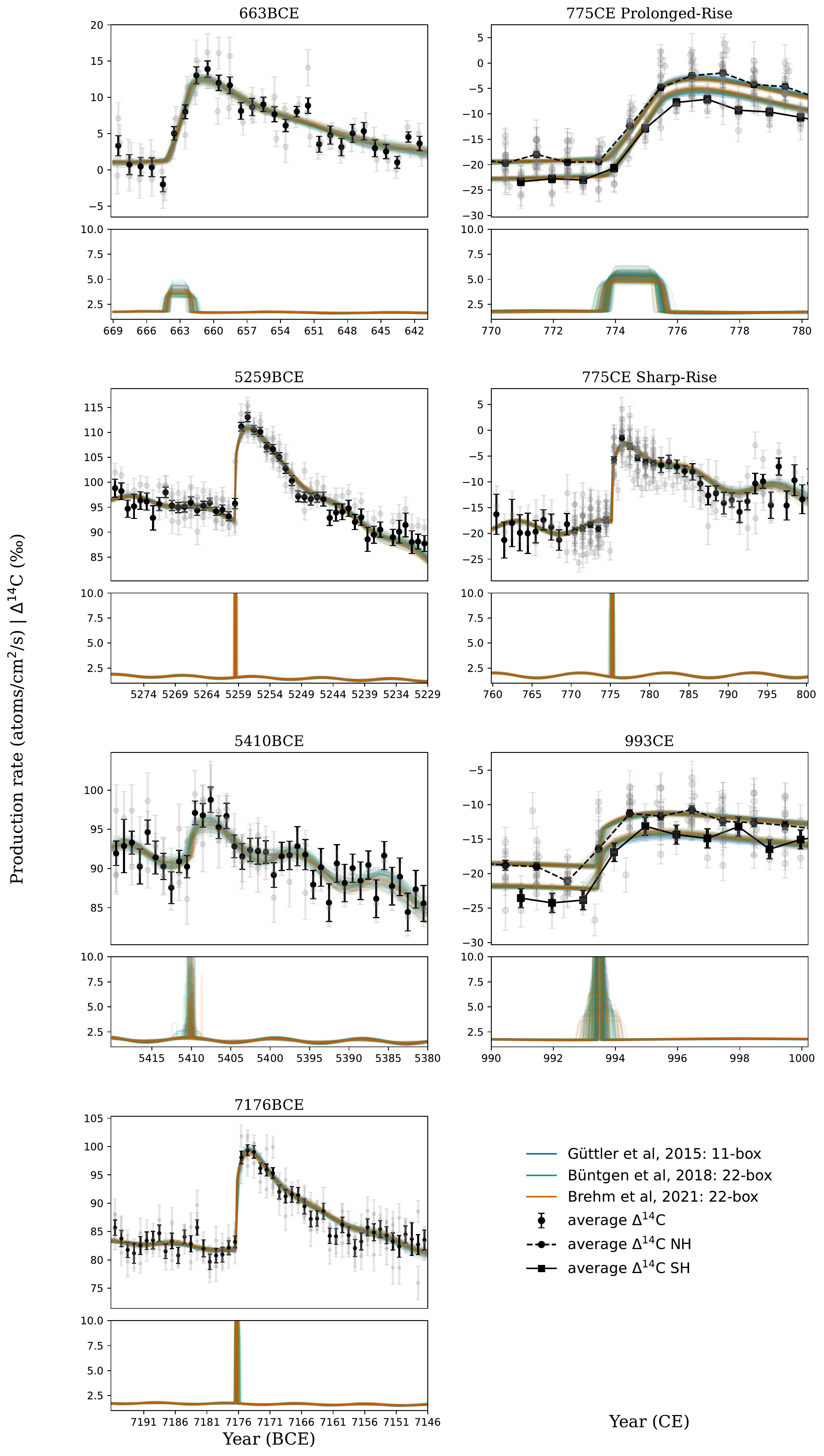}
    \caption{Results of \ac{mcmc} fitting of a parametric Miyake event model to all six known events. Each is presented in a pair of panels. Top: the tree-ring data (individual trees in grey, mean in black) overlaid with colour-coded curves drawn at random from \ac{mcmc} posterior samples for all three \ac{cbm}s; they are in excellent agreement with one another and with the data. Bottom: radiocarbon production rate models drawn from the corresponding \ac{mcmc} posterior samples, with the same colour bars. The 663\bce event and a subset of the 774\ce event are consistent only with a production spike taking longer than a year. The 774\ce event is presented split into subsets of data showing a prolonged rise, and a sudden rise, which are incompatible in our models and analysed separately. }
    \label{fig:fits-samples}
\end{figure}

Posterior ensembles of models overlaid on data, together with corresponding radiocarbon production histories, are displayed in Figure~\ref{fig:fits-samples}, and they show overall excellent agreement with data. Corner plots of the parameter posteriors are available in Supplementary Online Material. %, Figures~\ref{fig:corner_993} to~\ref{fig:corner_7176}.

% distribution of fits

\subsection{Nonparametric Retrieval of Production Rates}
\label{sec:nonparametric}

\begin{figure}
    \centering
    \includegraphics[scale=0.45]{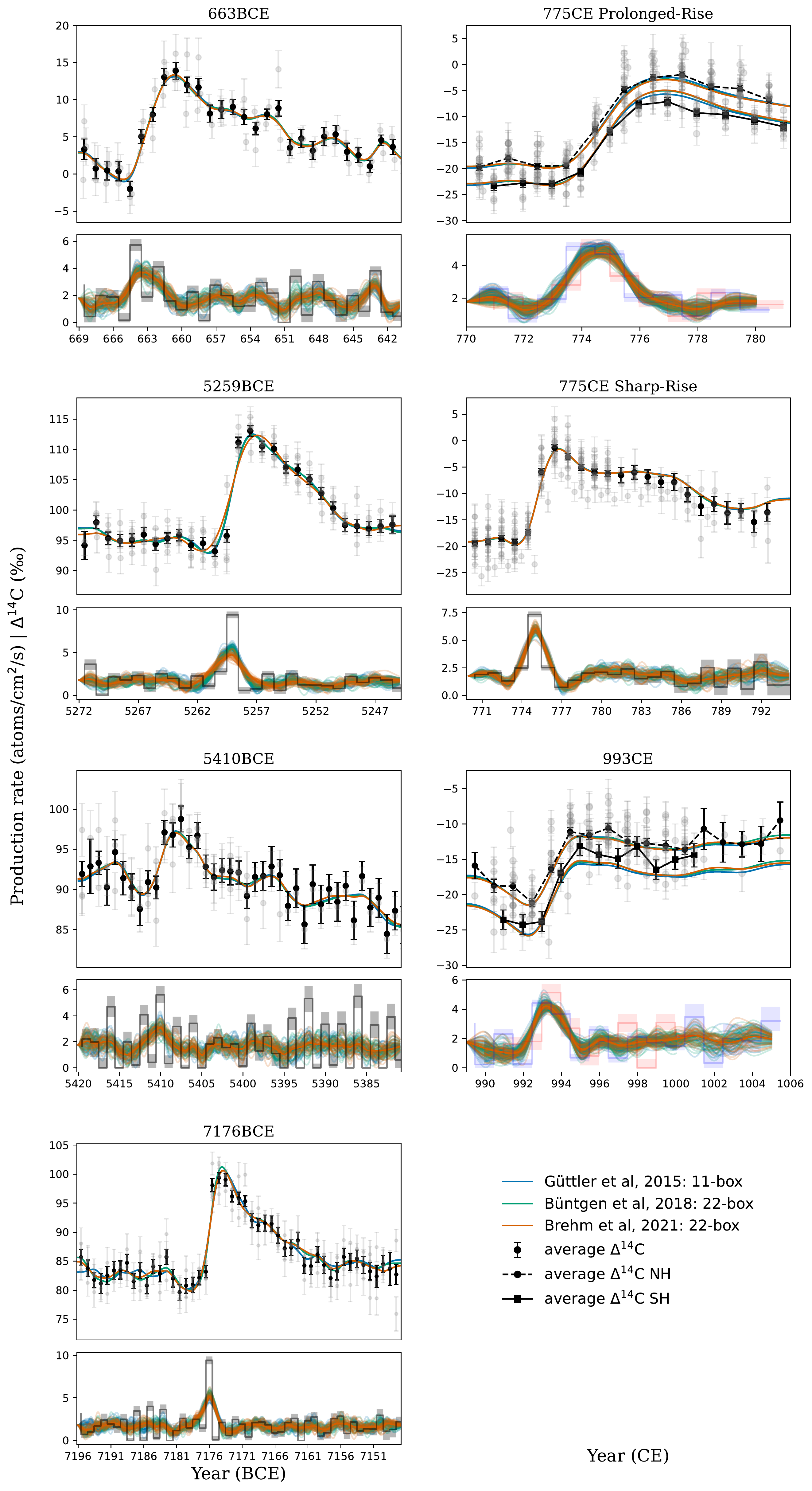}
    \caption{Results of \ac{mcmc} fitting of a nonparametric control-points radiocarbon production rate model to all six known events. Each is presented in a pair of panels. Top: the tree-ring data (individual trees in grey, mean in black) overlaid with colour-coded curves drawn at random from \ac{mcmc} posterior samples for all three \ac{cbm}s. Bottom: radiocarbon production rate models drawn from the corresponding \ac{mcmc} posterior samples, with the same colour bars; durations greater than a year are a necessary feature of the control points method and not strong evidence of long duration events. The 774\ce event is presented split into subsets of data showing separately a prolonged rise, and a sudden rise.}
    \label{fig:control-points}
\end{figure}

In addition to the parametric fits displayed above, we applied both the GP and inverse solver nonparametric retrievals to the same datasets, and visualize the output similarly in Figure~\ref{fig:control-points}. We again obtain a good fit to data, with the events occurring in the expected years, though now without the possibility of deconvolving structure at very short timescales. The GP and inverse solver produce results that are consistent with one another.

As an extension for future work, it is feasible to apply the inverse solver to the entire IntCal20 history, and use this as an initialization point for the reservoirs and production history of parametric fits at any particular point in time; this is a plausible strategy for making like-for-like comparison in absolute radiocarbon production between events occurring at times with different baseline production rates. We have elected not to do so here, to avoid introducing spurious transients in our sinusoidal production model, and without knowing a straightforward way to resolve this tension.

% non-parametric control-points

\section{Discussion}
\label{sec:discussion}

We find excellent agreement between the three carbon box models emulating \citet{guttler15}, \citet{brehm21}, \citet{buntgen18}, with a closer agreement between the latter two models which are partitioned into northern and southern hemispheres. In contrast, as noted by \citet{usoskin2013}, \citet{miyake17} has a different normalization\footnote{In \citet{miyakephd} it is explained that an equilibrium production rate is assumed over a $\pi R_\oplus^2$ Earth cross-sectional area, as opposed to a $4 \pi R_\oplus^2$ isotropic area, and this leads to a different assumption about equilibrium production rates.} and excludes the substantial carbon reservoir of the deep ocean, and we have excluded this from plots.

\subsection{Timing of 774\,\ce Event}
\label{sec:timing_774}
The 774\,\ce event occurs almost synchronously  across the range of species and locations involved. However, there is some variation in the rate at which the increase is expressed. Broadly there are two types of increases. About half of the data sets support a sharp rise - an anomalous jump in the data within one year - and the other half a more prolonged rise over 2-3 years. Furthermore, the latter group includes trees of the same species, in similar locations, measured at the same laboratories. 

A similar late rise is found in Japanese cedar for the 993\,\ce event by \citet{Miyake2022}, who interpret this as being affected by global atmospheric circulation patterns in different latitudinal Radiocarbon Zones, and an oceanic versus continental distinction. In contrast, in the ensemble of tree rings over 774\,\ce, prolonged and sharp rises are seen across these categories: there are trees showing both phenomenologies from Zones 0-2, continental or oceanic regions, different growth speeds and altitudes. It is therefore not completely clear what is the cause of this split in 774\,\ce phenomenologies. In future work there may be insights from global circulation models of the atmosphere, together with improved precision and sample size for tree-ring data over these events.

\subsection{Miyake Event Amplitude and Duration}

\begin{figure}
    \centering
    \includegraphics[scale=0.45]{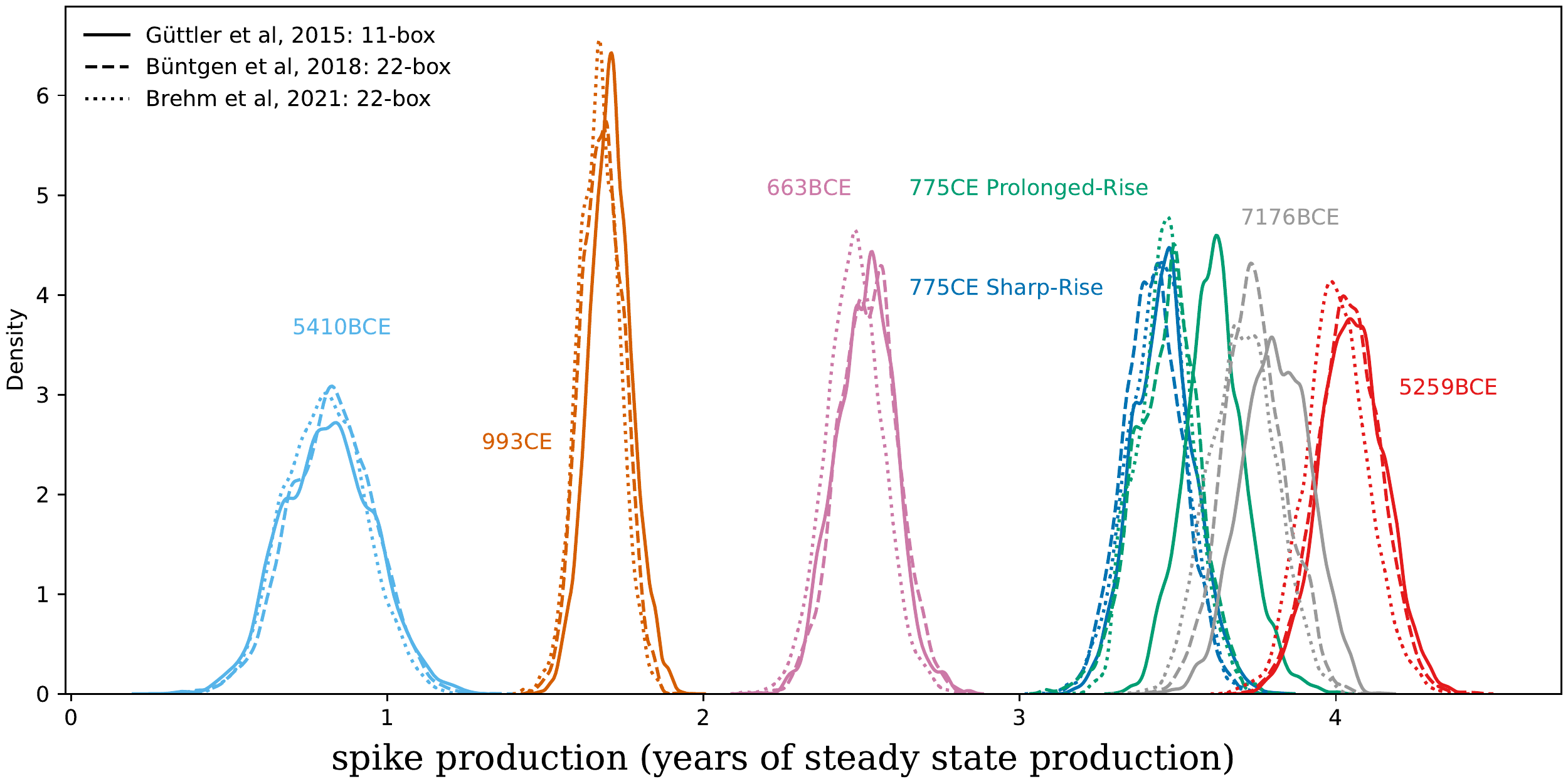}
    \caption{Marginal posterior probability distributions for the total radiocarbon production inferred for all six known Miyake events. This is calculated as the area under a spike, irrespective of its duration, minus the steady state, in units of equivalent years of steady state production. Different events are denoted by colour, and different \ac{cbm}s by solid \citep{guttler15}, dashed \citep{buntgen18}, and dotted \citep{brehm21} lines. The datasets showing a short or prolonged rise for 774\ce are shown separately.}
    \label{fig:spike-hist}
\end{figure}

In order to investigate the astrophysical origin of the Miyake events, it is of primary importance to determine their fundamental parameters - especially their size and duration. Posterior distributions of spike production relative to the steady state are displayed in Figure~\ref{fig:spike-hist}. 

Because we work with \dfc rather than absolute \fc, we report the integrated spike radiocarbon production in units of equivalent years of steady state production: i.e., a spike amplitude of 1 in these units indicates a total production of $1\,q_0\,\text{yr}$. In these units, the smallest event is 5410\,\bce, with a total production a shy of 1 $q_0\,\text{yr}$, followed by 993\,\ce at around 2 $q_0\,\text{yr}$ and 663\,\bce at around 2.5 $q_0\,\text{yr}$. The 775\,\ce and 7176\,\bce events are in excess of 3 $q_0\,\text{yr}$, and the largest of all is 5259\,\bce at around $4\,q_0\,\text{yr}$. 

We are intentionally wary of attempting a conversion of this to absolute kg\fc, which would bring in the more-uncertain $q_0$ at the time of each event, or of correcting this for the geomagnetic field, as this requires assumptions about the origin and spectrum of particles. Nevertheless it is interesting that in these units the spike amplitudes are all of order unity --- as might be expected from a change of order unity to heliospheric shielding of Galactic cosmic rays, for a duration of order 1 year.

\begin{figure}
    \centering
    \includegraphics[scale=0.45]{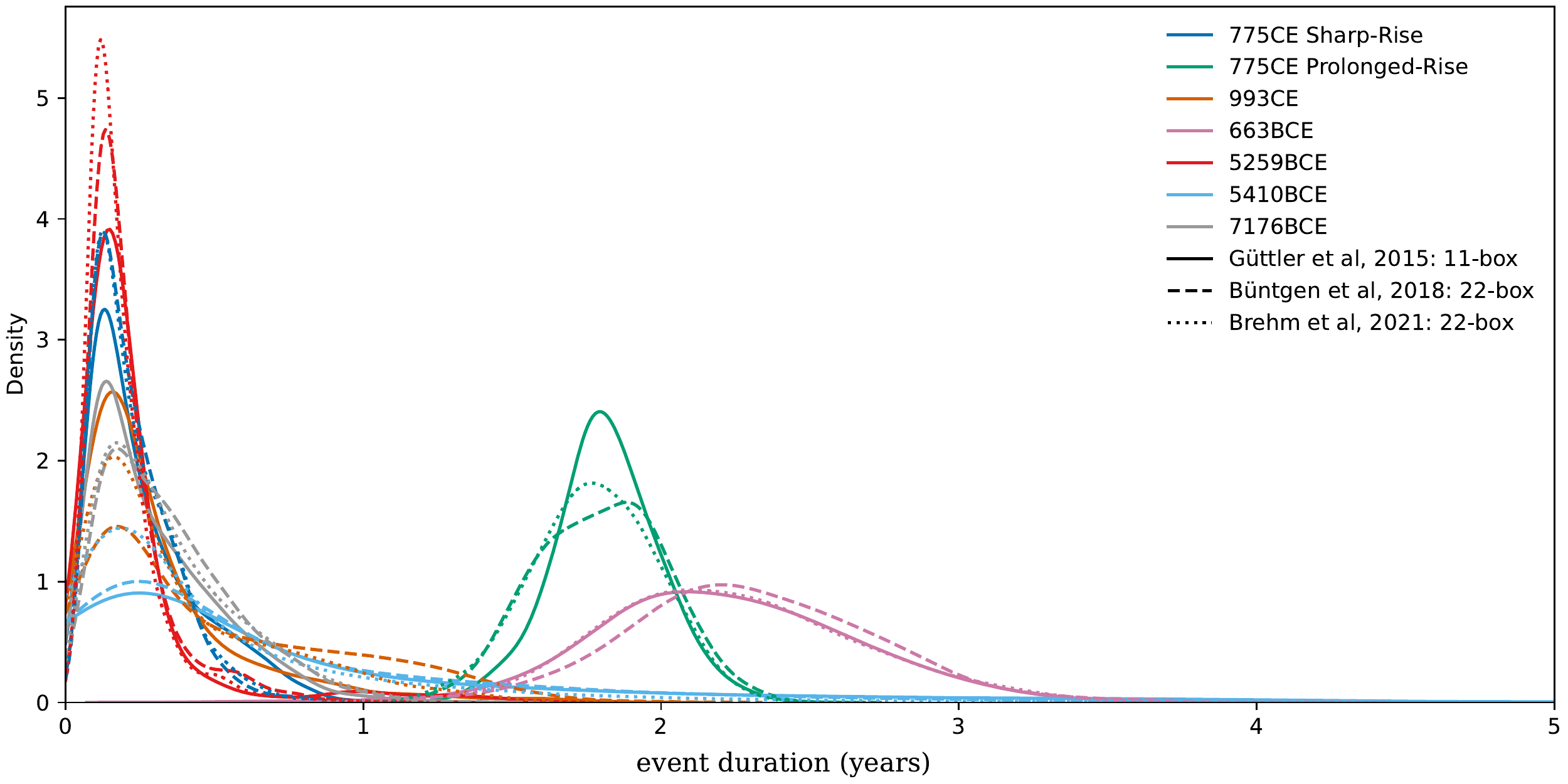}
    \caption{Marginal posterior probability distributions for the duration of all six known Miyake events. Different events are denoted by colour, and different \ac{cbm}s by solid \citep{guttler15}, dashed \citep{buntgen18}, and dotted \citep{brehm21} lines. The datasets showing a short or prolonged rise for 774\ce are shown separately.}
    \label{fig:spike-duration}
\end{figure}

The marginal posterior distributions for duration are displayed in Figure~\ref{fig:spike-duration}, showing that while 7176\,\bce, 5259\,\bce,  and a subset of 775\,\ce data are consistent with durations of $<1\,\text{yr}$, the duration of 5410\,\bce is very poorly constrained, a subset of 775\,\ce data indicate a duration of around 2 years (for these trees, $100\%$ of posterior samples have durations $>1$\,yr, and $\sim 15\% > 2$\,yr), and 663\,\bce has a duration of 2-3 years. There is a somewhat extended tail in the posterior for 993\,\ce, with 20\% of samples showing durations $>6$\,months, although only $\sim 4\% > 1$\,yr. These are all covariant with start date, as seen in the corner plots in the Supplementary Online Material: an early start and a long duration, or a late start and a brief duration, are both compatible with the data due to the 1-year sampling limitation. This is marginal evidence against a model of the Miyake events arising from a single short impulse; this can only be confirmed with multi-isotopic data, such as from ice cores where finer time sampling is achievable, and from a better understanding of the systematics induced by growth seasons and geography in tree-ring time series.

% density plot of spike production in steady state production

% density plot of spike duration

% density plot of spike timing 
\begin{figure}
    \centering
    \includegraphics[scale=0.45]{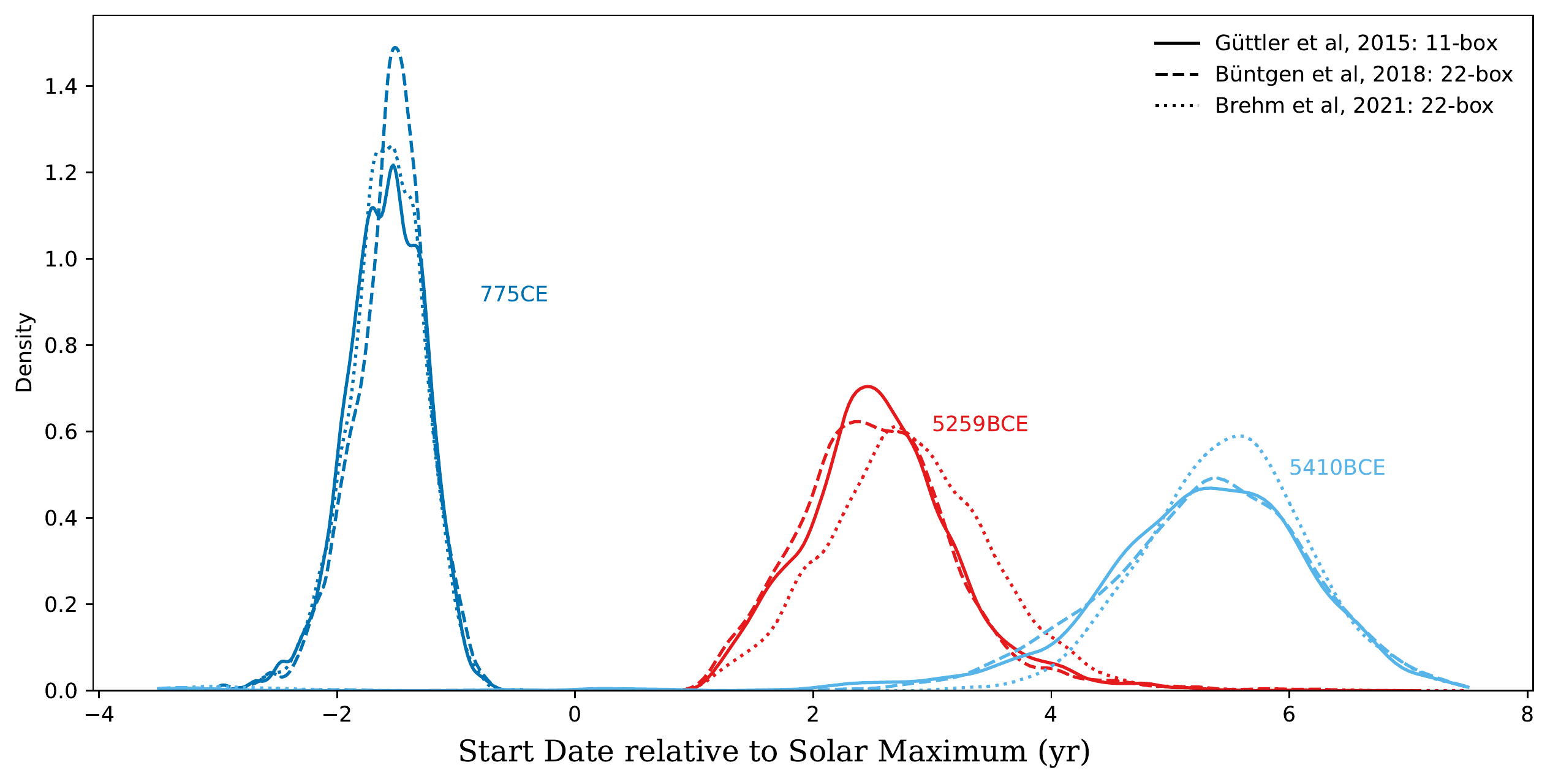}
    \caption{Marginal posterior probability distributions for the timing relative to the Solar Cycle of three Miyake events for which significant solar cycles are detected: 774\ce, 5259\bce and 5410\bce. Assuming the minimum of solar activity corresponds to a maximum of radiocarbon production rate, we find that 5259\bce and 5410\bce occur at or shortly before solar minima, while 774\ce occurs two years before maximum. Different events are denoted by colour, and different \ac{cbm}s by solid \citep{guttler15}, dashed \citep{buntgen18}, and dotted \citep{brehm21} lines.}
    \label{fig:solar-cycle}
\end{figure}

\subsection{Relation to the Solar Cycle}

High solar magnetic field strength gives rise to low \fc production, because the solar magnetic field shields the Earth from galactic cosmic rays. Therefore, we can define solar maxima to be the minima of the 11-year sinusoidal component of \fc production. Our Bayesian posteriors for solar cycle phase $\phi$ show no consistent pattern. Moreover, our histograms of the event timing, relative to solar cycle in Figure~\ref{fig:solar-cycle}, also show no obvious connection. We find that 5410\bce occurs at or shortly before solar minimum, 5259\ce a couple of years after, while 774\ce occurs two years before maximum.
The 993\ce event is more difficult to resolve. The increase seems to occur at a solar minimum  in some runs, but the analysis is currently lacking in sufficient data for any firm conclusions. As a result this event is excluded from Figure~\ref{fig:solar-cycle} but further data will undoubtedly improve on this outcomes. With regard to the 7176\bce event, \citet{Paleari2022} believe the \tbe evidence supports its occurence at a solar minimum. In contrast to that study as well as \citet{scifo19} and \citet{miyake21}, our findings show no clear relationship between the appearance of one of these events and the phase of the solar cycle, though with only three examples so far we cannot statistically reject any dependence.

\subsection{Dependence on Latitude}

\begin{figure}
    \centering
    \includegraphics[scale=0.6]{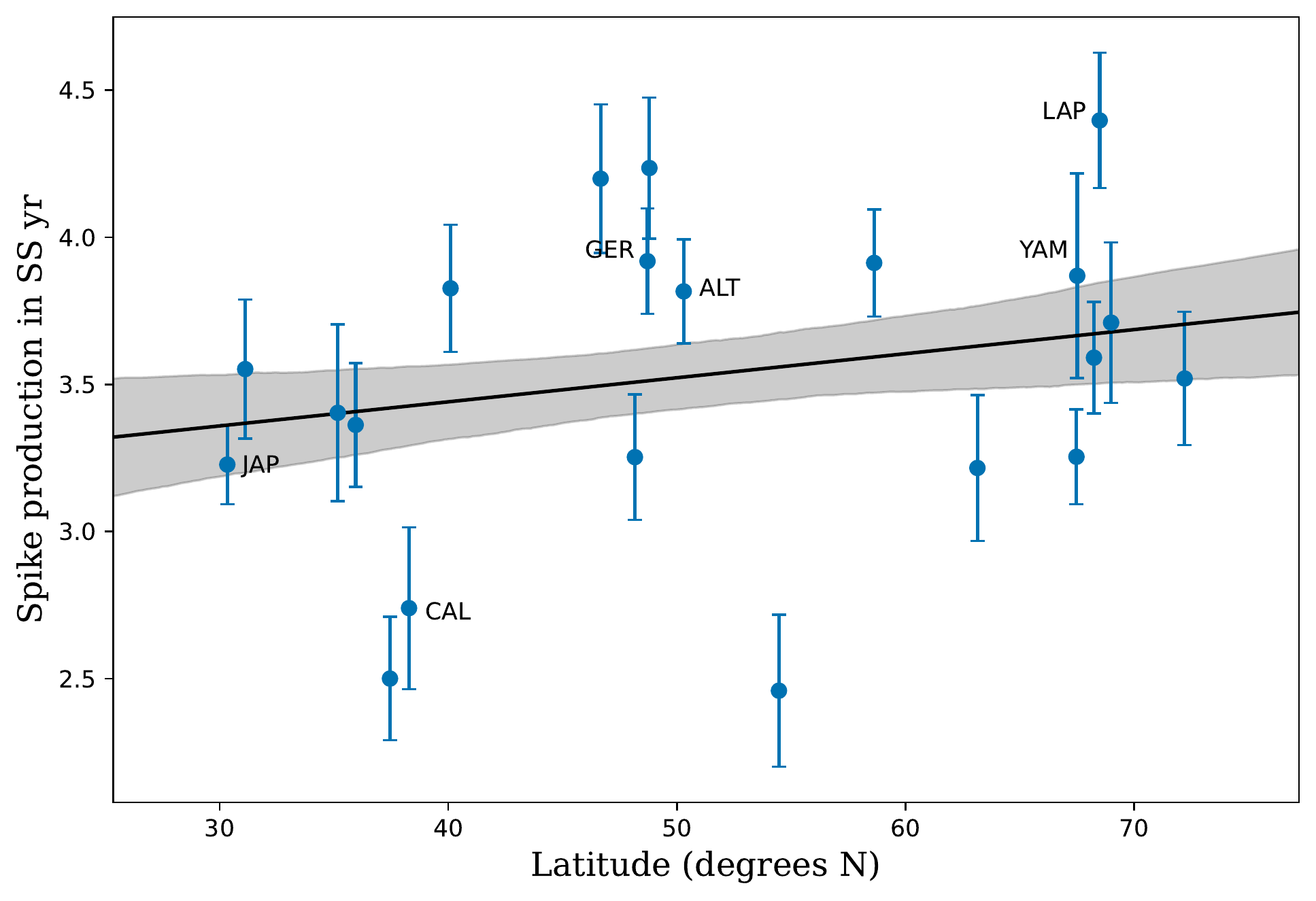}
    \caption{Scatter plot with 1\,$\sigma$ errorbars of 774\ce spike amplitude recorded in individual trees as a function of latitude, overlaid with best-fit line and shaded between \{16,84\} percentiles of posterior draws. While a trend with latitude has been used to support a solar origin \citep{Uusitalo2018}, with our larger sample of trees we find the slope to be $(9.8 \pm 8.8) \times 10^{-3}$ steady state years per degree north}
    % -- consistent with zero. 
    \label{fig:latitude}
 \end{figure}

In the case of an extreme solar event, a greater particle flux and therefore radiocarbon production is expected near the poles than the equator. Both \citet{buntgen18} and \citet{Uusitalo2018} claim that the amplitude of the events as recorded by northern hemisphere tree rings is increased closer to the North Pole. In order to examine this possible trend, we fit every tree individually for spike amplitude, timing and duration, while holding parameters of the solar cycle constant at the ensemble mean. We then used \texttt{emcee} to fit a line and infer parameter uncertainties, including an additional term for underestimated error bars. Our outputs provide no convincing evidence for this effect. With our larger sample of trees, we find the slope to be $(9.8 \pm 8.8) \times 10^{-3}$ steady state years per degree north, with a 13\% probability the slope is less than zero.
In the main, this possible latitudinal trend largely goes away because of the scatter observed in the array of data available from mid-latitudes, see Figure~\ref{fig:latitude}.

% \subsection{Uncertainty of the Solar Cycle}
% Qingyuan's resampling + bandpass filtering: can we even see the solar cycle in much of these data?

% \input{prolonged_table}

\section{Conclusion}

In this work we have combined fast \ac{cbm} models with modern Bayesian inference tools, and applied them to the ensemble of existing data on Miyake events. From the posterior parameter distributions we infer, we find no clear relation in timing to the solar cycle, or in amplitude to latitude as has previously been claimed; and we find some evidence of extended duration not only in 663\,\bce, but also in 775 and to some extent in 993\,\ce. This can be interpreted either as a real astrophysical nonzero duration, or as a noise floor on time resolution owing to the growth conditions and biology of the trees, model uncertainties in the preindustrial carbon cycle, and/or to atmospheric dynamics not captured by the carbon box model. In order to resolve this question, in future work we will want to obtain larger samples of high-precision, annual-cadence tree-rings over these well-studied events; multiradionuclide time series, including subannually-resolved \tcl and \tbe from ice-cores; and systematically compare \ac{cbm} implementations to global circulation models that accurately capture the latitudinal and stratosphere-troposphere exchange of radiocarbon. 

If the measured extended durations are reproduced, and are owing to biological or atmospheric processes, these will impose a precision floor of order $\sim 1\,\text{yr}$ on radiocarbon dating with these Miyake events \citep[such as done for L'Anse aux Meadows  by][]{Kuitems2021}. On the other hand, if the prolonged radionuclide production has an astrophysical origin, this will be hard to reconcile with an impulsive production model of one large energetic particle burst, whether of solar energetic particles or from a stellar remnant. In light of this, we recommend that it is important to obtain improved multiradionuclide data across the 5480\,\bce decade-long radiocarbon rise, and the three-year 663\,\bce event, as they may form a continuum in duration with the other shorter radiation bursts. 

There is very significant scope to improve open-source software for carbon isotope analysis. \ticktack can emulate the parameters of a range of existing \ac{cbm}s: future work would also systematically compare the different \ac{cbm} parameters to one another and to data in a range of contexts, including varying the stratosphere-to-troposphere production coefficients.  \ticktack is also extensible to connect to other inference tools: because the solver is implemented in \textsc{Jax}, it would be straightforward to implmement a Hamiltonian Monte Carlo sampler for very large or complex models using the probabilistic programming language \texttt{numpyro} \citep{Phan2019}, including a more sophisticated treatment of priors. 
A project beyond \ticktack would solve for multiple isotopes, and include effects of atmospheric dynamics and geochemistry. In the long term, it would be worth applying \ac{mcmc} to more complex models for a variety of applications, including inferring the parameters of the preindustrial carbon cycle directly from radiocarbon data, or inferring growth seasons and timing of different trees. We expect that there will be many applications for fast, differentiable carbon cycle models connected to modern Bayesian frameworks across geo- and astro-physics.

\vskip6pt

\enlargethispage{20pt}

% \ethics{Insert ethics text here.}

\dataccess{In the interests of open science, we have made the \ticktack code available under an MIT open source license at \href{https://github.com/sharmallama/ticktack}{github.com/sharmallama/ticktack}, with documentation provided at \href{https://sharmallama.github.io/ticktack}{sharmallama.github.io/ticktack}. The Snakemake workflow used to analyse the data is available at \href{https://github.com/qingyuanzhang3/radiocarbon_workflow}{github.com/qingyuanzhang3/radiocarbon\_workflow}. We encourage and welcome other scientists to replicate, apply, and extend our work.}

\aucontribute{QZ and US contributed equally to code development and data analysis and prepared all figures. JD developed and tested \ac{ode} solvers. AS, MK, and MWD provided advice on radiocarbon dynamics and processing. MJO contributed to solar physics content. BJSP conceived and supervised the project, code, and writing. All authors contributed to the text.}

\competing{The authors declare no competing interests.}

\funding{This research was supported by the UQ Winter and Summer Research Scholarships, and the inaugural UQ Fellowship of the Big Questions Institute.}

\ack{We would like to thank Andrew Smith, David Fink, Quan Hua, and the anonymous referees for their helpful comments on this manuscript.\\ \\
We would like to acknowledge the traditional owners of the land on which the University of Queensland is situated, the Turrbal and Jagera people. We pay respects to their Ancestors and descendants, who continue cultural and spiritual connections to Country.\\ \\
This research made use of the \textsc{IPython} package \citep{ipython}; Snakemake \citep{snakemake}; \textsc{NumPy} \citep{numpy}; \textsc{matplotlib} \citep{matplotlib}; \textsc{SciPy} \citep{scipy}; Google \textsc{Jax} \citep{jax}; \texttt{ChainConsumer} \citep{chainconsumer}; \texttt{emcee} \citep{emcee}; \textsc{JaxNS} \citep{jaxns}; and we emulate the models of \citet{guttler15}, \citet{miyake17}, \citet{buntgen18}, and \citet{brehm21}. }

% \disclaimer{Insert disclaimer text here.}

%%%%%%%%%% Insert bibliography here %%%%%%%%%%%%%%

\bibliography{ms} % if your bibtex file is called example.bib

\appendix
\newpage

\section*{Supplementary Online Material: Carbon Box Model Implementation}

% \begin{figure}
% \centering
% \end{figure}

Many closed-source implementations of \ac{cbm}s exist, with varying levels of publicly-available documentation. In the interests of reproducibility, we document here key assumptions of the \ticktack model.

% what ticktack is for vs what it is not

\begin{itemize}

    \item \textbf{Carbon-14 Half Life}: 5700 years  \citep{godwin62,kutschera2019}.

    \item \textbf{Growth Seasons:} When comparing to annual tree-ring data, the \ac{cbm} outputs are not taken as point samples but rather integrated over a top-hat window over the domain of the tree's growth season, which can be specified arbitrarily. In this Paper, trees are all taken to have a 6-month growing season centred on mid-summer in their hemisphere, but in future work this limitation can be relaxed. Early and late wood is treated computationally and in figures as having growth seasons in the first and second halves of this six month period.

    \item \textbf{Dating Convention:} For radiocarbon dating, the commonly used convention for reporting dates is from \citet{schulman1956}. According to this convention, the calendar years often do not correspond to the dates provided in radiocarbon data, rather the start of the year is defined as the start of the growth season for each tree. For example, the 770 data point associated with a tree of growth season from April to July refers to the \dfc content from 769 April to 770 April, not 769 January to 770 January. Points in figures have an abscissa value centred on the middle of their assumed growth season, for example 774.0 for a southern tree ring but 774.5 for a northern one. 
    
    \item \textbf{\dfc in 22-box models:} The \citet{buntgen18} and \citet{brehm21} models are partitioned into northern and southern hemispheres. When loading data from a tree from the northern or southern hemisphere, and when converting from $^{14}C$ to \dfc, the corresponding hemisphere's troposphere values are used automatically.
    
    \item \textbf{Fluxes are Balanced} : For each reservoir, radiocarbon is a negligible proportion of the total quantity of carbon. The $^{12}$C reservoirs are assumed to be in a steady state, which means that the $^{12}$C fluxes ($F_{C, i \rightarrow j}$) into and out of a reservoir are balanced:
    \[\frac{d N_{C,i}}{dt} = \sum_{j=1}^{J} F_{C, i \rightarrow j} + \sum_{j=1}^{J} F_{C, j \rightarrow i} = 0\]
    \item \textbf{Radiocarbon Flux is proportionate}: no fractionation of $^{14}$C relative to $^{12}$C is included in the model, and $^{13}$C is ignored. The absolute flux of $^{14}$C is therefore
    \begin{equation}
        N_{^{14}\text{C},j}/N_{^{12}\text{C},j} \cdot F_{C,j\rightarrow{i}}
    \end{equation}
    
    \item \textbf{Unit Conversion}: Two systems of units can be used for both production rate and flux: production rates can be specified in atoms/cm$^2$/s or kg/yr, fluxes in Gt/yr or 1/yr, and these are automatically converted appropriately. We assume where production rates are in atoms/cm$^2$/s that the surface area of the Earth is a full $4\pi R_\oplus^2$ and not the cross-sectional $\pi R_\oplus^2$ used in \citet{miyake12}, and adopt a baseline production rate of $q_0 = 1.76$\,atoms/cm$^2$/s following \citet{brehm21}.
    
    \item \textbf{No production of $^{14}C$ occurs outside the atmosphere} : All production of $^{14}C$ takes places in the reservoirs specified in the model and as such, the sum of production coefficients across the reservoirs is normalized to 1. We do not simulate in situ production of radiocarbon in trees.
    
    \item \textbf{Long-term production history:} In order to avoid a spurious transient effect in our simulations from mismatch between a flat previous production history and a sinusoidal production over the window of interest, we initialize each computation with a `burn-in' for N years (typically $\sim 1000$ at coarse 1-year sampling, and then the reservoir values at year \textit{i} are used to initialize the starting values of the boxes for the first epoch of data modeling.
    
    \item{ \textbf{Equilibrium Carbon Reservoirs:} There are two types of equilibration provided by \ticktack: either with \textbf{fixed production}, where production rate is held constant and the steady state solution is determined by matrix inversion of the flow matrix; or with a \textbf{fixed reservoir}, where a single reservoir is held fixed, and the production rate is found to reach this equilibrium. We use the inverse matrix approach above to find reservoirs as a function of production rate, and use \textsc{Jax} gradient descent and a BFGS optimiser to find the corresponding production rate.}
    \item \textbf{Time-Invariant Dynamics:} all dynamics are time-invariant and no atmospheric dynamics or Suess effect are included.
    % \item \textbf{Miyake model}: one arrow = both direction fluxes of equal magnitude
    \item \textbf{Solar Cycle Duration}: the solar cycle duration is fixed to 11.0\,yr in all parametric models.
    \item \textbf{Normalisation}: for each dataset, we subtract the mean of the first 4 \dfc data points  from all \dfc data, so that the Miyake event does not affect the \dfc baseline measurement.
    \item \textbf{ODE Solvers:} 
    The \ac{cbm} is a linear system of \ac{ode}s driven by an arbitrary production term $Q(t)$. Because the production term can rise on timescales as short as days, but the oceans respond on timescales of millennia, it can be a very stiff \ac{ode}. In existing models \cite{brehm21}, \cite{buntgen18}, \cite{guttler15}, \cite{miyake12}, a first-order finite difference method is used to numerically approximate the solution. In \ticktack, we solve this \ac{ode} system with higher-order numerical methods. 
    
    In order to find an effective model, we tested the accuracy on the analytic impulse response solution of a normalized 11-box \ac{cbm} \cite{guttler15}. 
    
    % So I think I should mention that they were not tested against the impulse response but rather each other. Alternatively just remove this section 
    Using the \texttt{DifferentialEquations.jl} package \citep{diffeqjl} in Julia \citep{julia2017}, we tested a variety of higher-order methods, finding that a 5th-order Dormand-Prince algorithm \citep[DP5;][]{dp5} had similar time performance to a first-order solver but much higher accuracy. In \ticktack we use the default \textsc{Jax} DP5 implementation. The accuracy of the first-order finite difference method was very sensitive to the time sampling of the simulation. When the time sample was in the range of $~10\to 100$ samples per the median absolute error was in the range $~1\times 10^{-4}$, compared with $~1\times 10^{-8}$ for the DP5 algorithm with the same sampling. A typical likelihood call for the 11-box model on a single dataset for the 774\ce event would take {5--10\,ms} to run.
    
\end{itemize}

\subsection*{Box Coefficients}
\begin{figure}[H]
\centering
\includegraphics{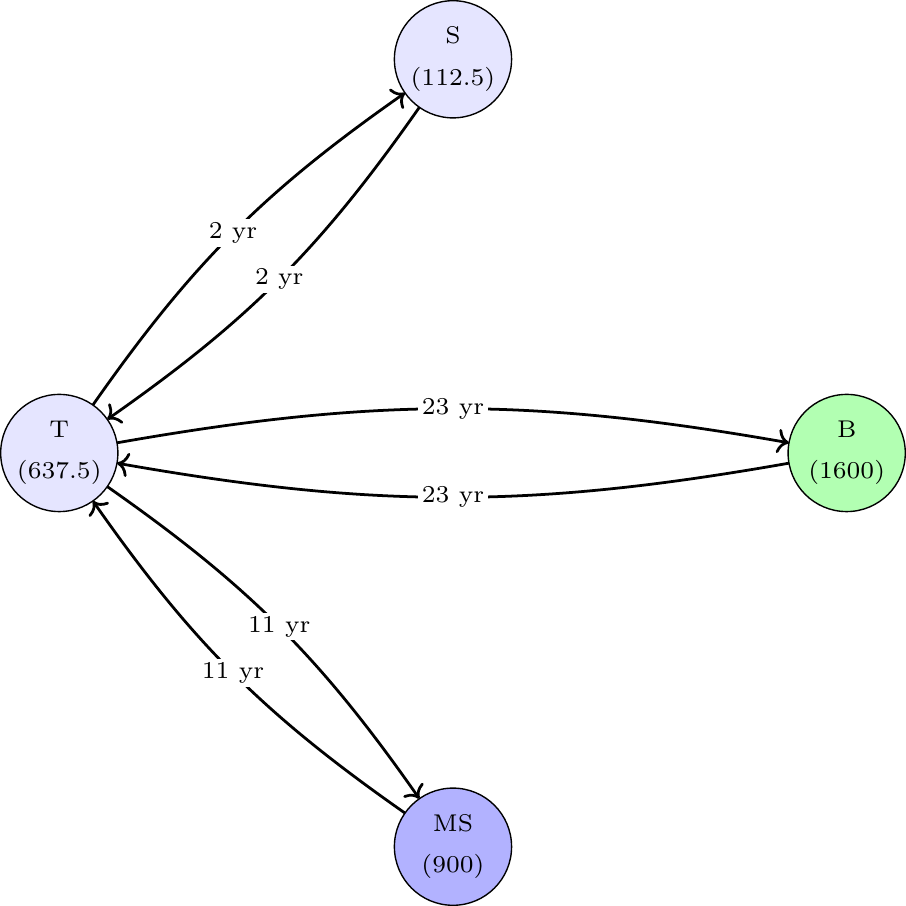}
\vspace{0.5cm}
\includegraphics{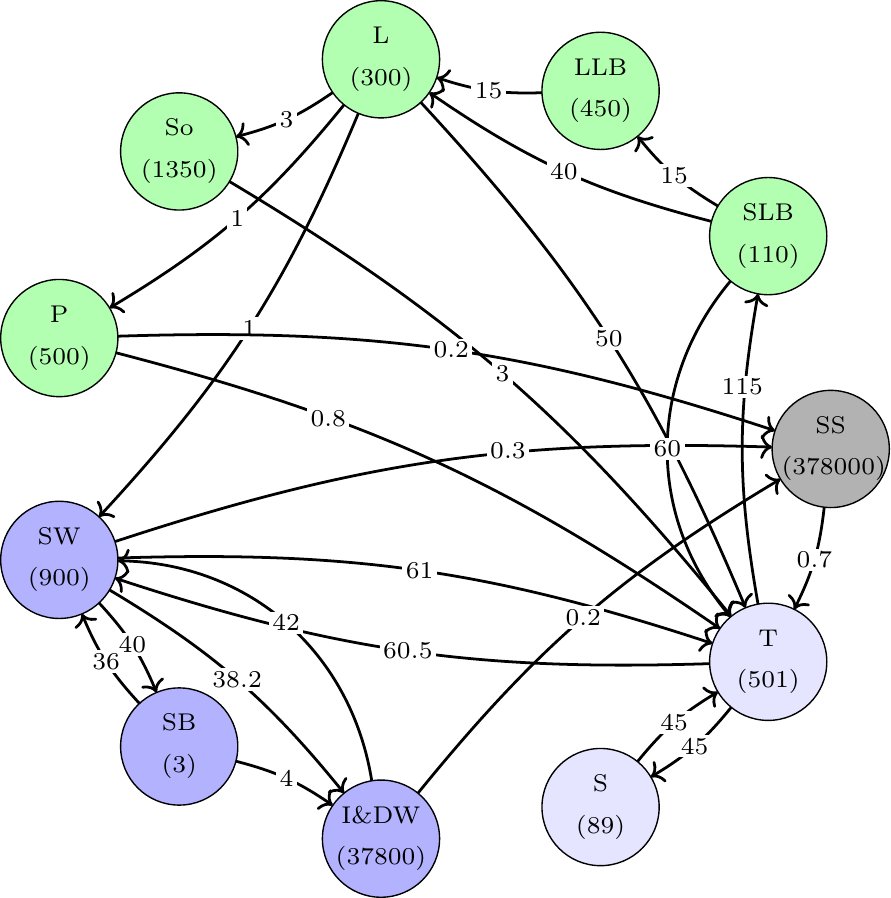}
\caption{Carbon box models following \citet{miyake17} (top: 4-box, flux units in residence time) and \citet{guttler15} (bottom: 11-box, flux units in Gt/yr). All reservoir contents are given in Gt underneath their respective reservoirs. }
\label{fig:miyake-guttler}
\end{figure}

\begin{figure}[H]
\centering
\includegraphics{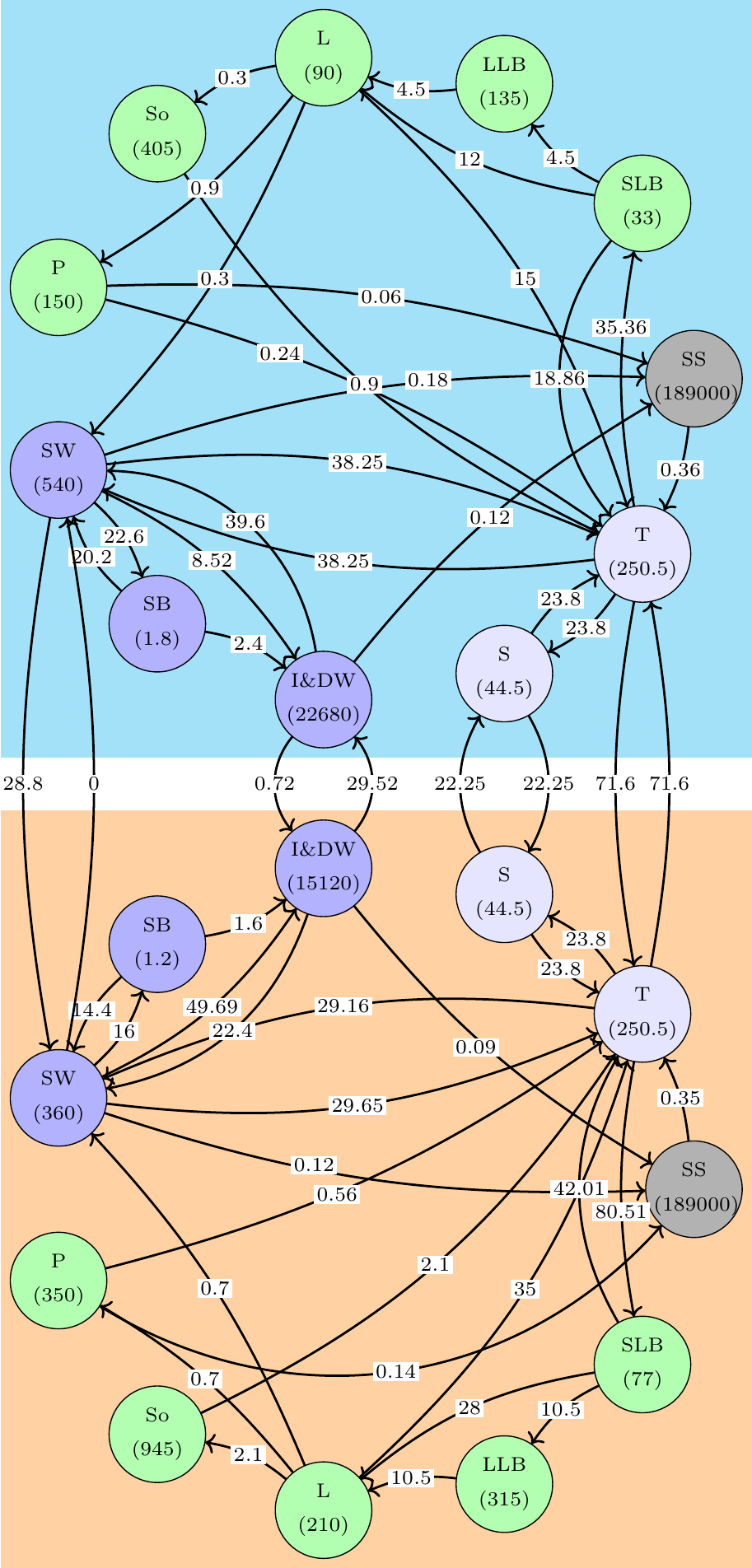}
\caption{22-box carbon box model following \citet{buntgen18}. Flux units are Gt/yr. Reservoir contents are given in Gt underneath their respective reservoirs.}
\label{fig:buntgen_model}
\end{figure}

\begin{figure}[H]
\centering
\includegraphics{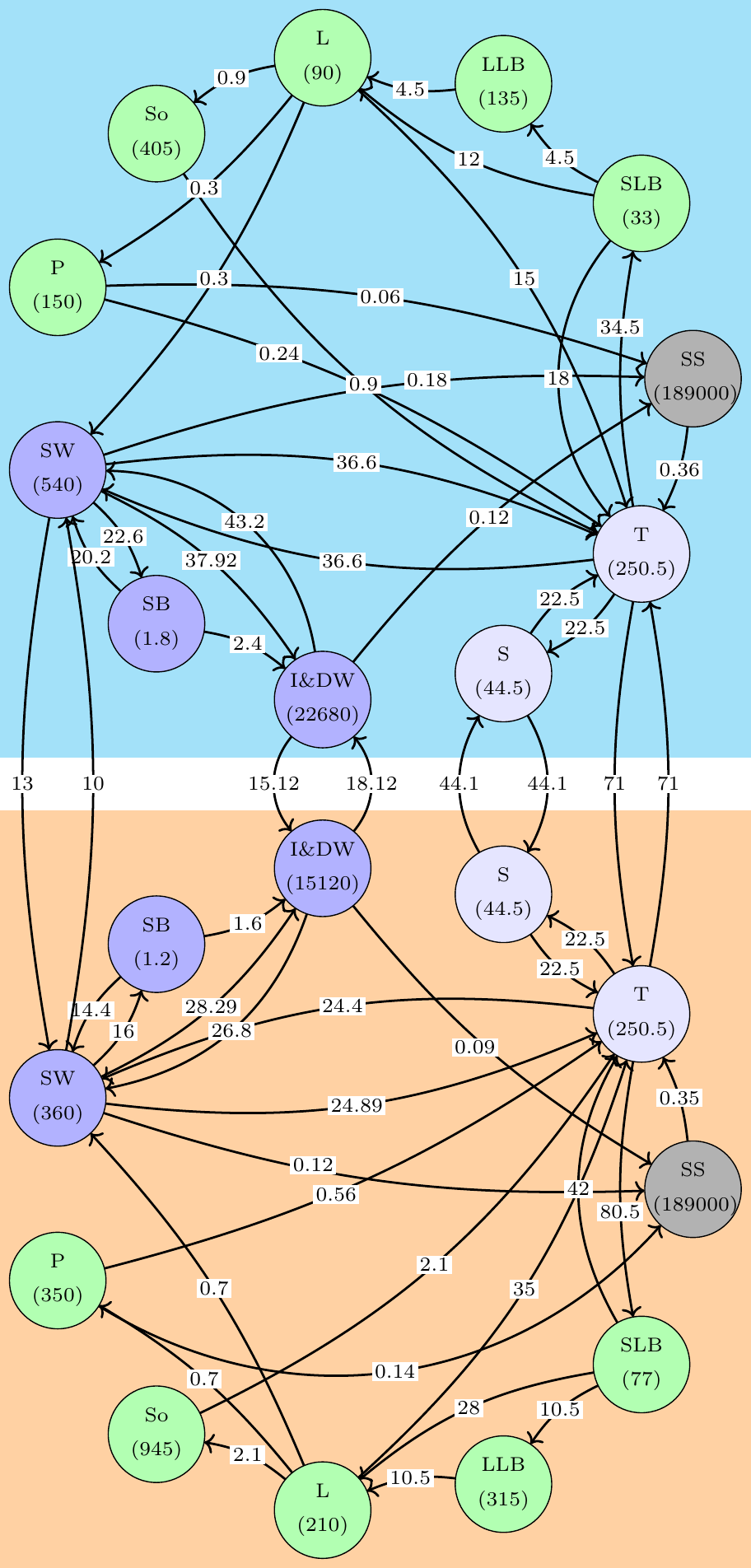}
\caption{22-box carbon box model following \citet{brehm21}. Flux units are Gt/yr. Reservoir contents are given in Gt underneath their respective reservoirs.}
\label{fig:brehm_model}
\end{figure}

In Figures ~\ref{fig:miyake-guttler} to~\ref{fig:brehm_model}, we display the topologies, reservoirs and flow coefficients of the carbon box models emulated in this paper, first presented in \citet{guttler15}, \citet{miyake17}, \citet{buntgen18}, and \citet{brehm21}. The latter two models are split into northern and southern hemispheres, with identical labels in both.

The abbreviations for the reservoir labels are:

\begin{description}
\item[T] Troposphere
\item[S] Stratosphere
\item[B] Biosphere
\item[SLB] Short-Lived Biota
\item[LLB] Long-Lived Biota
\item[L] Litter
\item[So] Soil
\item[P] Peat
\item[MS] Marine Surface
\item[SW] Surface Water
\item[SB] Surface Biota
\item[I\&DW] Intermediate and Deep Water
\item[SS] Sedimentary Sink
\end{description}

\newpage

\newpage

\section{Event Posterior Parameter Distributions}

In Supplementary Online Material, we display posterior sample distributions from \ac{mcmc} as corner plots, rendered using ChainConsumer \citep{chainconsumer}. Machine-readable tables of these \ac{mcmc} chains are available from \href{https://github.com/qingyuanzhang3/radiocarbon_workflow}{github.com/qingyuanzhang3/radiocarbon\_workflow}.

\begin{figure}[H]
    \centering
    \includegraphics[scale=0.45]{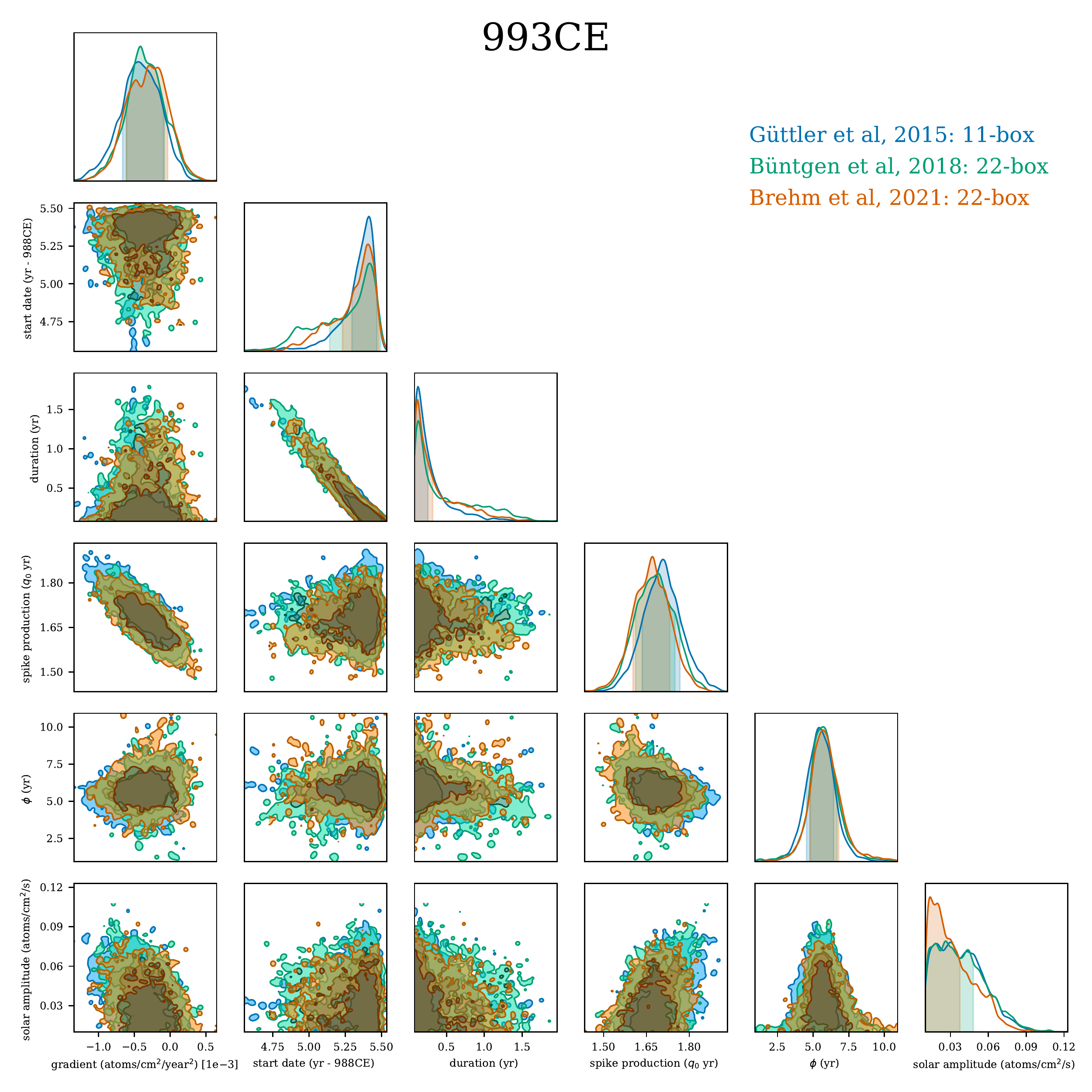}
    \caption{Corner plot of posterior parameter samples for the 993\ce event, inferring a gradient, start date, duration, total spike production, and phase and amplitude of the 11-year solar cycle. We simultaneously display posteriors from the \ac{cbm}s of \citet{guttler15} in blue, \citet{buntgen18} in green, and \citet{brehm21} in orange. \ac{mcmc} chains produced using \texttt{emcee} \citep{emcee} and rendered using ChainConsumer \citep{chainconsumer}.}
    \label{fig:corner_993}
\end{figure}
\newpage

\begin{figure}[H]
    \centering
    \includegraphics[scale=0.45]{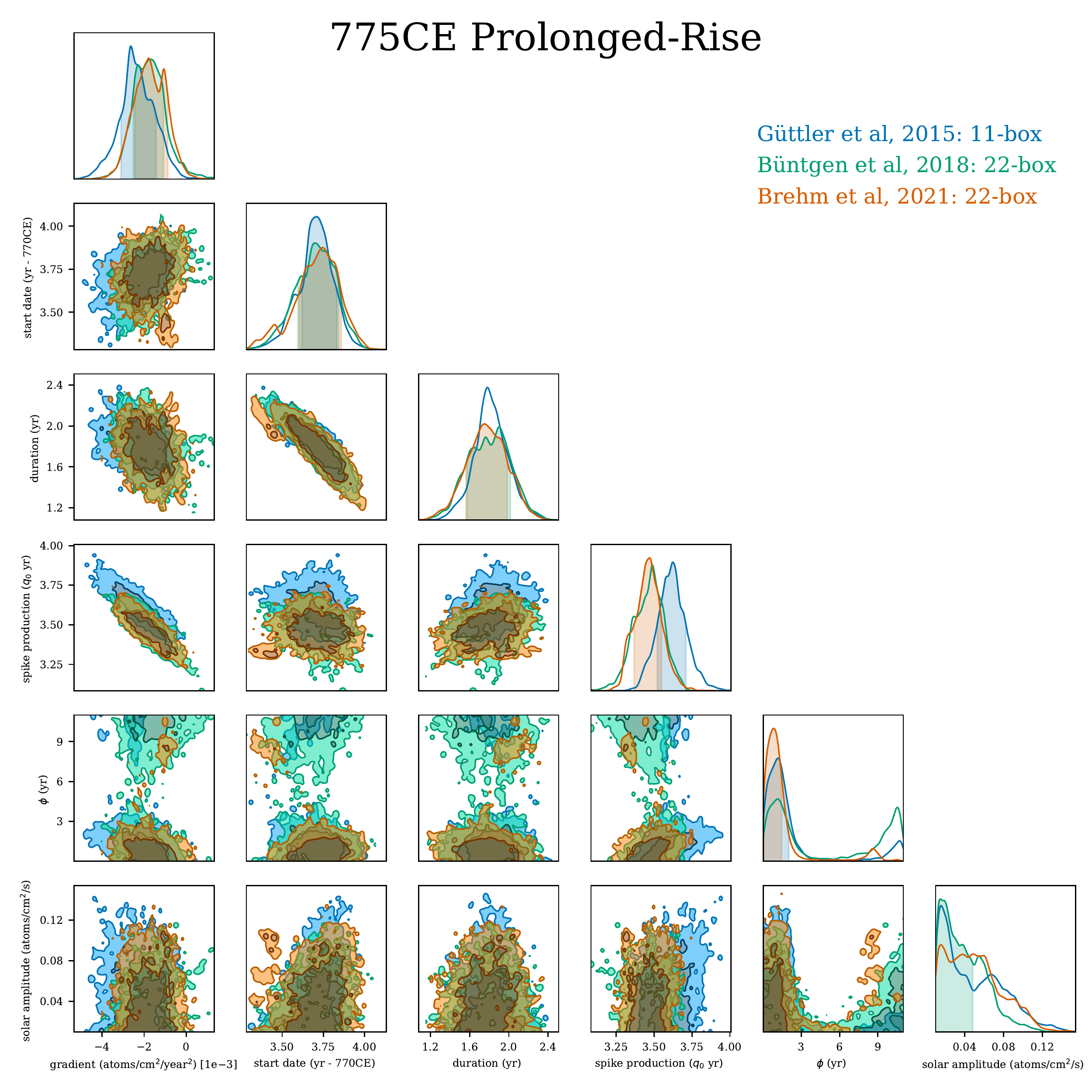}
    \caption{Corner plot of posterior parameter samples for the 774\ce event using only samples showing a prolonged rise, inferring a gradient, start date, duration, total spike production, and phase and amplitude of the 11-year solar cycle. We simultaneously display posteriors from the \ac{cbm}s of \citet{guttler15} in blue, \citet{buntgen18} in green, and \citet{brehm21} in orange. \ac{mcmc} chains produced using \texttt{emcee} \citep{emcee} and rendered using ChainConsumer \citep{chainconsumer}.}
    \label{fig:corner_774long}
\end{figure}
\newpage

\begin{figure}[H]
    \centering
    \includegraphics[scale=0.45]{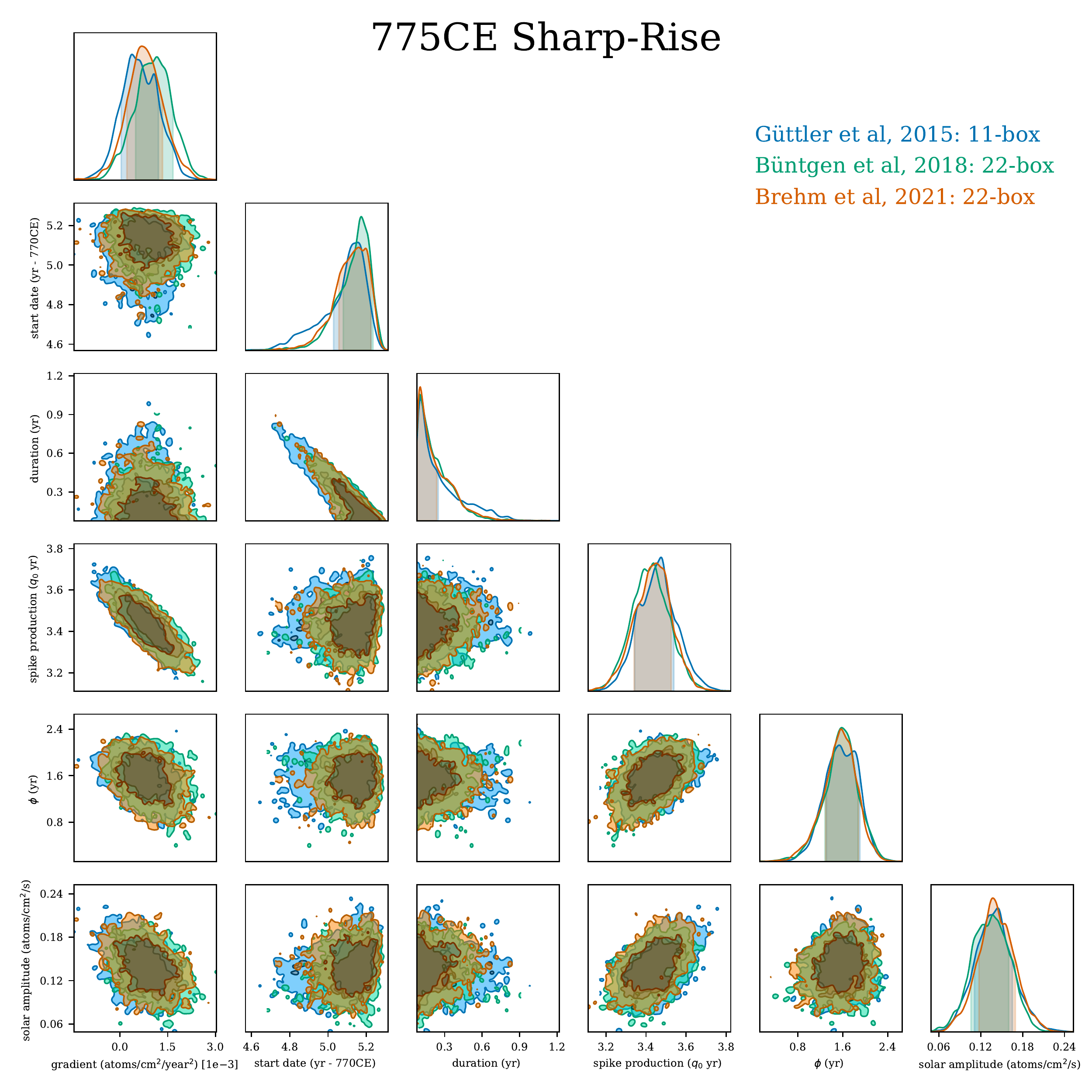}
    \caption{Corner plot of posterior parameter samples for the 774\ce event using only samples showing a sharp rise, inferring a gradient, start date, duration, total spike production, and phase and amplitude of the 11-year solar cycle. We simultaneously display posteriors from the \ac{cbm}s of \citet{guttler15} in blue, \citet{buntgen18} in green, and \citet{brehm21} in orange. \ac{mcmc} chains produced using \texttt{emcee} \citep{emcee} and rendered using ChainConsumer \citep{chainconsumer}.}
    \label{fig:corner_774short}
\end{figure}
\newpage

\begin{figure}[H]
    \centering
    \includegraphics[scale=0.45]{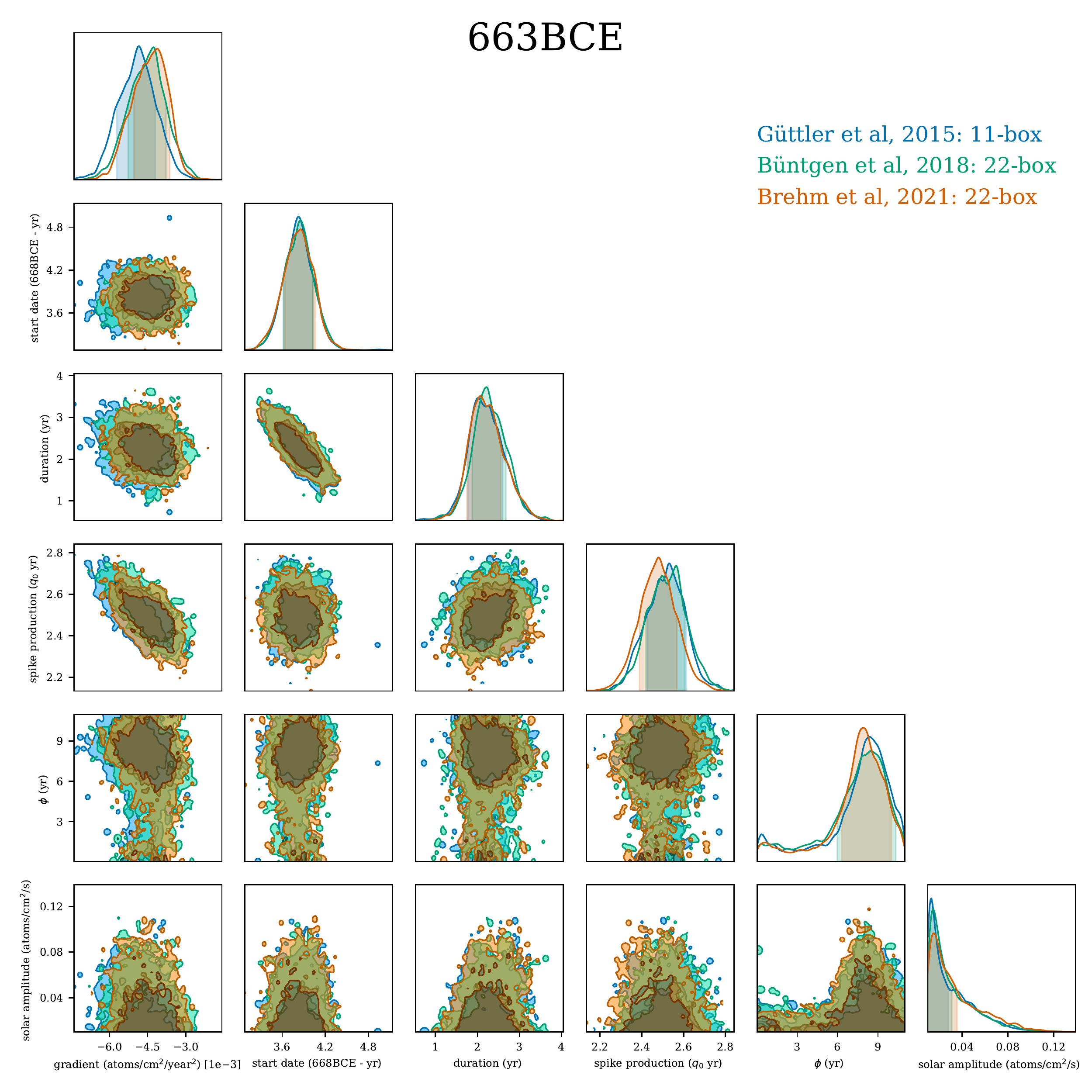}
    \caption{Corner plot of posterior parameter samples for the 663\bce event, inferring a gradient, start date, duration, total spike production, and phase and amplitude of the 11-year solar cycle. We simultaneously display posteriors from the \ac{cbm}s of \citet{guttler15} in blue, \citet{buntgen18} in green, and \citet{brehm21} in orange. \ac{mcmc} chains produced using \texttt{emcee} \citep{emcee} and rendered using ChainConsumer \citep{chainconsumer}.}
    \label{fig:corner_663}
\end{figure}
\newpage

\begin{figure}[H]
    \centering
    \includegraphics[scale=0.45]{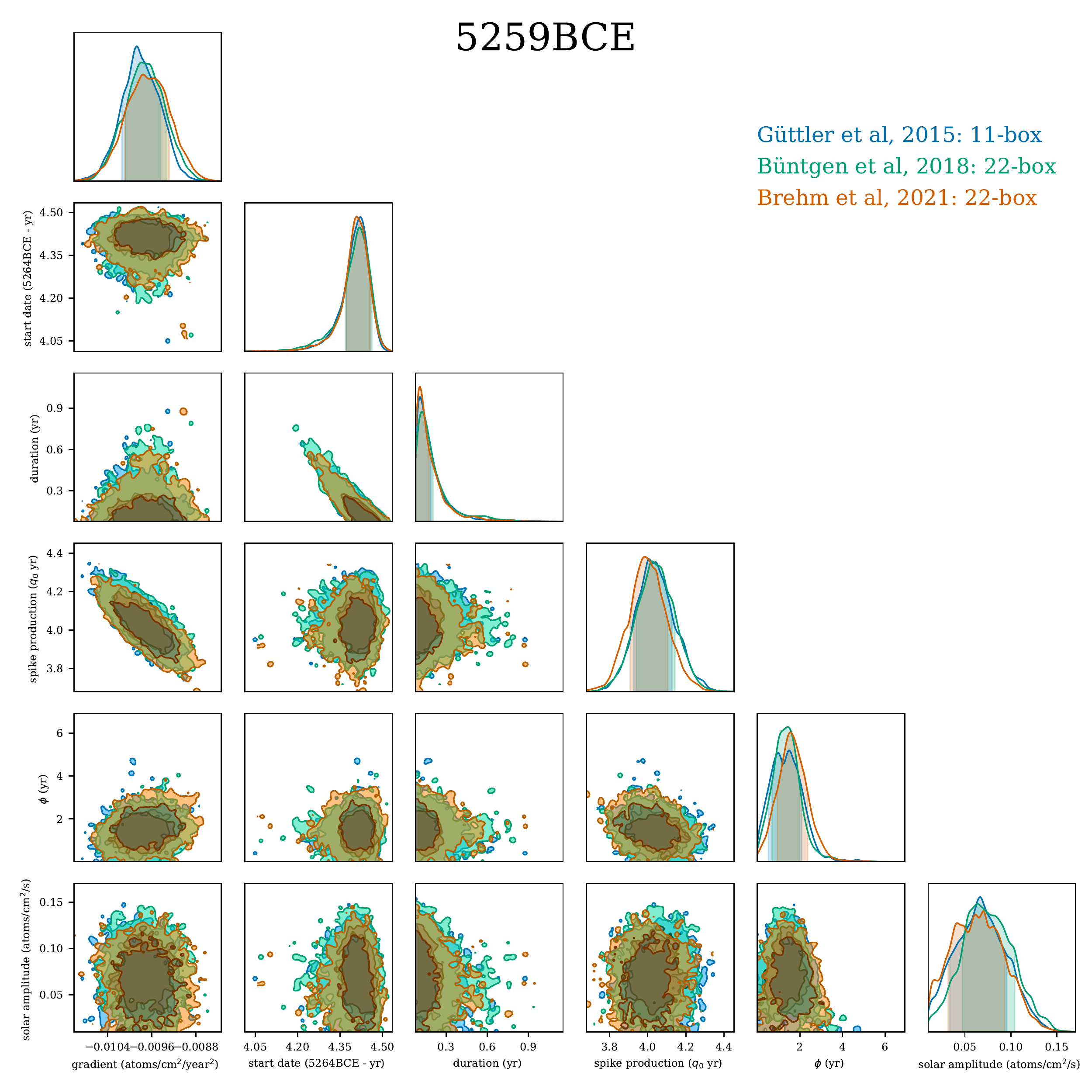}
    \caption{Corner plot of posterior parameter samples for the 5259\bce event, inferring a gradient, start date, duration, total spike production, and phase and amplitude of the 11-year solar cycle. We simultaneously display posteriors from the \ac{cbm}s of \citet{guttler15} in blue, \citet{buntgen18} in green, and \citet{brehm21} in orange. \ac{mcmc} chains produced using \texttt{emcee} \citep{emcee} and rendered using ChainConsumer \citep{chainconsumer}.}
    \label{fig:corner_5259}
\end{figure}
\newpage

\begin{figure}[H]
    \centering
    \includegraphics[scale=0.45]{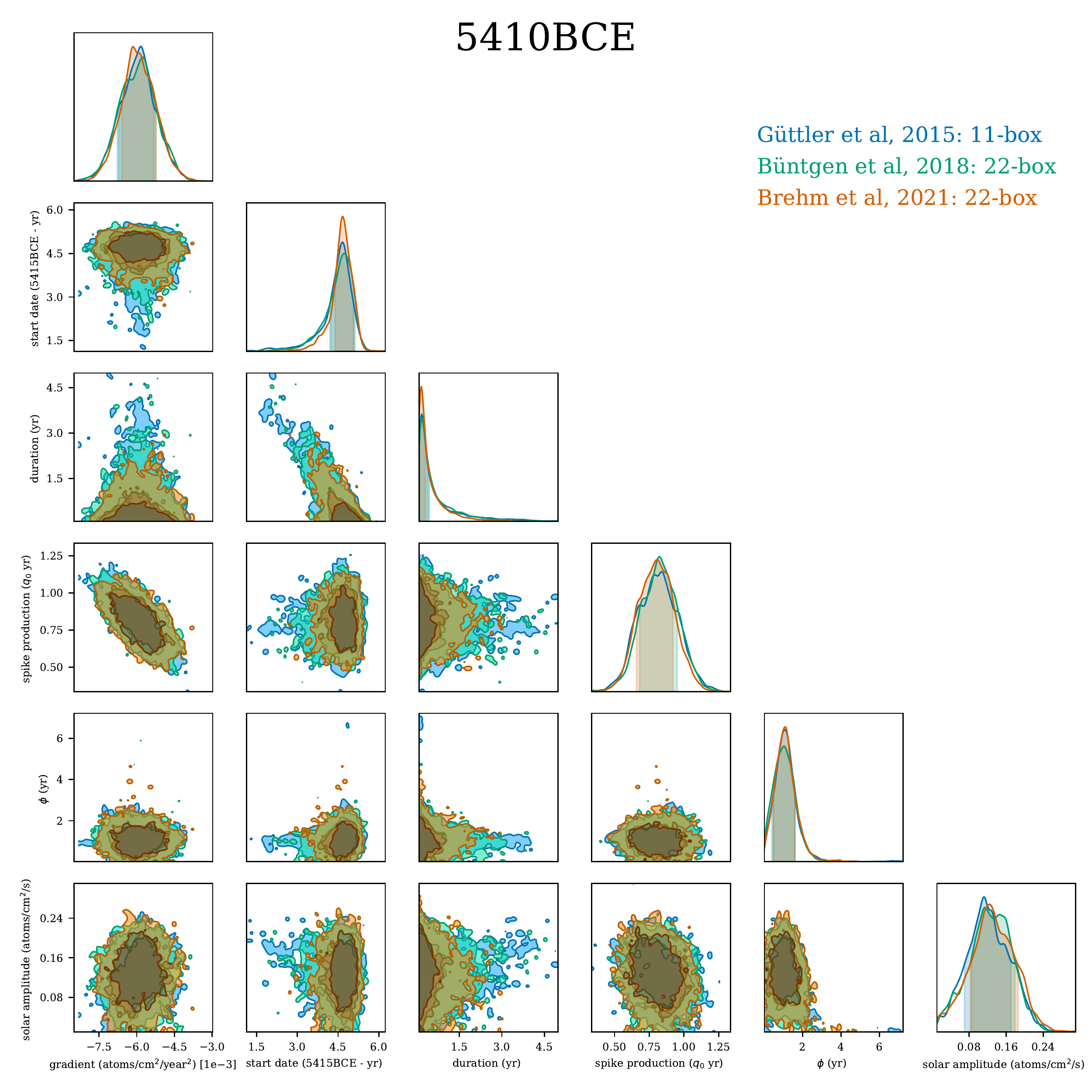}
    \caption{Corner plot of posterior parameter samples for the 5410\bce event, inferring a gradient, start date, duration, total spike production, and phase and amplitude of the 11-year solar cycle. We simultaneously display posteriors from the \ac{cbm}s of \citet{guttler15} in blue, \citet{buntgen18} in green, and \citet{brehm21} in orange. \ac{mcmc} chains produced using \texttt{emcee} \citep{emcee} and rendered using ChainConsumer \citep{chainconsumer}.}
    \label{fig:corner_5410}
\end{figure}
\newpage

\begin{figure}[H]
    \centering
    \includegraphics[scale=0.45]{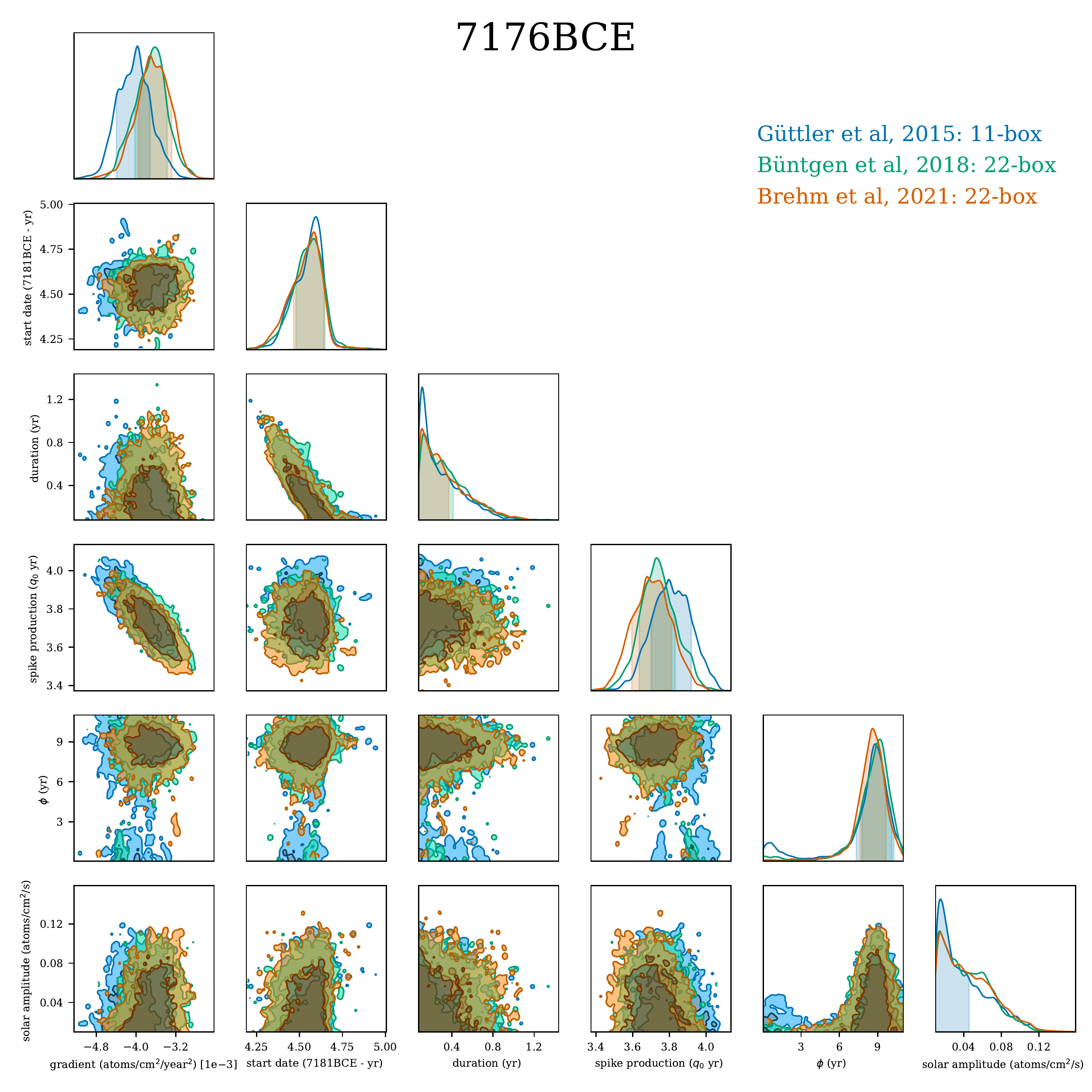}
    \caption{Corner plot of posterior parameter samples for the 7176\bce event, inferring a gradient, start date, duration, total spike production, and phase and amplitude of the 11-year solar cycle. We simultaneously display posteriors from the \ac{cbm}s of \citet{guttler15} in blue, \citet{buntgen18} in green, and \citet{brehm21} in orange. \ac{mcmc} chains produced using \texttt{emcee} \citep{emcee} and rendered using ChainConsumer \citep{chainconsumer}.}
    \label{fig:corner_7176}
\end{figure}
\newpage

\end{document}